\documentstyle[prb,aps]{revtex}
\def\rr{{{\bf r}}}
\def\rrp{{{\bf r^\prime}}}
\def\nup{{{n_\uparrow}}}
\def\ndo{{{n_\downarrow}}}

\begin{document}

\title{ Fragmentation of positively charged metal clusters in stabilized
jellium model with self-compression \footnote{This work is dedicated to the
memory of my mother, Gohar and the 68$^{th}$ birthday of my father, Bahram.}}

\author{M. Payami }
\address{Center for Theoretical Physics and Mathematics,
Atomic Energy Organization of Iran,\\ P.~O.~Box 11365-8486, Tehran, Iran}

\date{\today}
\maketitle
\begin{abstract}
Using the stabilized jellium model with self-compression, we have calculated
the dissociation energies and the barrier heights for the binary fragmentation
of charged silver clusters. At each step of calculations, we have used the
relaxed-state sizes and energies of the clusters. The results for the doubly
charged
Ag clusters predict a critical size, at which evaporation dominates the
fission, in good agreement with the experiment. Comparing the dissociation
energies and the fission barrier heights with the experimental ones, we
conclude that in the experiments the fragmentation occurs
before the full structural relaxation expected after the ionization of the
cluster. In the decays of Ag$_N^{4+}$ clusters, the results predict that
the charge-symmetric
fission processes are dominant for smaller clusters, and the charge-asymmetric
fission processes become dominant for sufficiently larger clusters.
\end{abstract}
\newpage

\section{Introduction}
\label{sec1}

The fragmentation of ionic metal clusters\cite{Naher,Uzi99} as well as other
 properties of metallic clusters have been extensively studied using
the jellium model (JM).\cite{ekardt,knight,brack93} In this model, the
discrete ions are replaced by a uniform positive charge background of density
$n=3/4\pi r_s^3$ in which $r_s$ is the bulk value of the Wigner-Seitz (WS)
radius of the valence electrons of the metal. The simplest geometry for the
positive background is spherical which is appropriate for closed-shell clusters
or large enough clusters in which the Jahn-Teller \cite{Jahn} deformation has
negligible contribution. However, using this simple spherical JM,
 a lot of information on the properties of metal clusters has been obtained.
A refined version of the JM, the stabilized\cite{pertran,Shore} jellium model
(SJM), which was introduced by Perdew {\it et. al.} in 1990, has improved some
drawbacks\cite{lang70,ashlang67} of the JM (For a recent review on SJM, see
 Ref.[\ref{Kiejna99}]). In recent years,
the SJM has been used to predict the properties of bulk metals
\cite{pertran,Perdew2001}, metal surfaces\cite{Fiolhais92,Kiejna99,Sarria},
 metal clusters\cite{Perdew93,Seidl,PayamiJPC} and metallic
voids.\cite{Ziesche} The fragmentation of charged
metal clusters has been also studied\cite{Vieira95,Vieira98}
by Vieira {\it et. al.} using the SJM. However, since the surface effects
have a large contribution in the energetics and sizes of small
clusters, and also since in a fission process the competition between the
surface tension and
coulomb repulsion leads to the existence of a barrier, a more sophesticated
use\cite{Perdew93,PayamiJPC} of the SJM is needed to predict
the correct energetics of the clusters and the barrier heights (BH) in the
study of the fragmentation processes.
This method, which is called SJM with self-compression (SJM-SC), has been used
to predict the equilibrium sizes and energies of charged\cite{Braj96}
or spin-polarized\cite{PayamiJPC} metal clusters as well as the calculation of
chemical potentials of metallic clusters.\cite{Kiejna96} The SJM-SC has been
also used by Sarria {\it et. al.}\cite{Sarria} to calculate the surface
energies and the work functions of metals.
In contrast to the JM and the SJM in which the $r_s$ value is borrowed from the
bulk system, in the
SJM-SC, the density parameter $r_s$ of the jellium sphere is assumed to be a
free parameter which can be adjusted in such a way that
a cluster
with a given number of electrons and specific electronic configuration
achieves its equilibrium state.
The SJM-SC calculations on neutral metal clusters\cite{Perdew93,PayamiJPC}
has shown that the
equilibrium $r_s$ value of the jellium sphere is less than the bulk value and
tends to its bulk value for infinitely large cluster. This
phenomenon is called self-compression which is due to the dominant effect of
surface tension in small metal clusters. However, it has been shown
that\cite{Braj96}
charging a small metal cluster can result in an equilibrium $r_s$ value which
is larger than the bulk value. This effect is called self-expansion. The
self-expansion has been also predicted for highly polarized metal
clusters\cite{PayamiJPC,Payami99}. These two effects have different origins. In
the former, the
repulsive coulomb force dominates the surface tension whereas, in the latter,
the Pauli force is responsible for the self-expansion.

In this work, using SJM-SC, we have
studied the binary decay processes of positively charged Ag clusters containing
up to 100 atoms in all possible channels.
We have considered the following possible decay processes for singly ionized
 Ag clusters
\begin{equation}
{\rm Ag}_N^{1+}\to {\rm Ag}_{N-p}^{1+} + {\rm Ag}_p^0,
\;\;\;\;\;\;\;\;p=1,2,\cdots,N-2.
\label{eq1}
\end{equation}
For doubly charged clusters, the decays can proceed via two
different processes. The first one is the evaporation process
\begin{equation}
{\rm Ag}_N^{2+}\to {\rm Ag}_{N-p}^{2+} + {\rm Ag}_p^0,
\;\;\;\;\;\;\;\;p=1,2,\cdots,N-3
\label{eq2}
\end{equation}
and the second one is fission into two charged products
\begin{equation}
{\rm Ag}_N^{2+}\to {\rm Ag}_{N-p}^{1+} + {\rm Ag}_p^{1+},
\;\;\;\;\;\;\;\;p=2,3,\cdots,[N/2]
\label{eq3}
\end{equation}

In general, for the
binary decay of $Z$-ply charged ($Z$ is a positive integer) cluster, we have
\begin{equation}
{\rm Ag}_N^{Z}\to {\rm Ag}_{N-p}^{Z-z_1} + {\rm Ag}_p^{z_1},
\;\;\;\;\;\;\;\;z_1=0,1,\cdots,[Z/2];\;\; p=z_1+1,\cdots,N-Z+z_1-1.
\label{decaygen}
\end{equation}
For an even value of $Z$ with $z_1=Z/2$, the range of $p$ reduces to
$p=z_1+1,\cdots,[N/2]$. The processes for which $z_1=0$ ( i.e., one of the
fragments is neutral), are called evaporation processes and others ( both
fragments are charged) are fission processes.
In evaporation processes, the negativity of the difference between total
energies before and after fragmentation is sufficient to have a spontaneous
decay. However, in fission processes a negative value for the difference
energy is not sufficient for the fission of the parent cluster. This is
because, the competition between the short-range
surface tension and the long-range repulsive coulomb force may give rise to a
fission barrier ( i.e., one should supply energy to overcome the barrier).

The organization of this paper is as follows.
In section \ref{sec2} we explain the method of calculating the total energies
and fission barriers.
To obtain the total energy of a given cluster, we solve
the self-consistent Kohn-Sham (KS) equations\cite{KohnSham}
in the density functional theory\cite{Kohn64} (DFT) with local spin density
approximation (LSDA) for the exchange-correlation (XC) functional.
To calculate the fission barrier, we use the two-touching-spheres model for the
saddle configuration.\cite{Naher}
In section \ref{sec3}, we
discuss the results, and finally, we conclude this work in section \ref{sec4}.

\section{Calculational Scheme}
\label{sec2}
\subsection{Total energy of a cluster}
In the context of the SJM, the average energy per valence electron in the
bulk
with density parameter $r_s$ and polarization $\zeta$ is given
by\cite{Payami98}

\begin{equation}
\varepsilon(r_s,\zeta,r_c)=t_s(r_s,\zeta)+\varepsilon_{xc}(r_s,\zeta)+\bar
w_R(r_s,r_c)+\varepsilon_{\rm M}(r_s),
\label{eq7}
\end{equation}

where

\begin{equation}
t_s(r_s,\zeta)=\frac{c_k}{r_s^2} \left[(1+\zeta)^{5/3}+(1-\zeta)^{5/3}\right]
\label{eq8}
\end{equation}

\begin{equation}
\varepsilon_{xc}(r_s,\zeta)=\frac{c_x}{r_s}
\left[(1+\zeta)^{4/3}+(1-\zeta)^{4/3}\right]+\varepsilon_c(r_s,\zeta)
\label{eq9}
\end{equation}

\begin{equation}
c_k=\frac{3}{10} \left(\frac{9\pi}{4}\right)^{2/3}
;\;\;\;\;c_x=\frac{3}{4}\left(\frac{9}{4\pi^2}\right)^{1/3}.
\label{eq10}
\end{equation}

All equations throughout this paper are expressed in Rydberg atomic units. Here
$t_s$ and $\varepsilon_{xc}$ are the mean noninteracting kinetic energy and
the exchange-correlation
energy per particle, respectively. For $\varepsilon_c$ we use the Perdew-Wang
parametrization.\cite{perwan}
For a $z$-valent metal the average Madelung energy, $\varepsilon_{\rm M}$, is
defined as
$\varepsilon_{\rm M}=-9z/5r_0$, in which $r_0$ is the radius of the WS sphere,
$r_0=z^{1/3}r_s$.
In Eq.(\ref{eq7}), $\zeta=(n_\uparrow-n_\downarrow)/(n_\uparrow+n_\downarrow)$
in which $n_\uparrow$ and $n_\downarrow$ are the spin densities of the
homogeneous system with total
density $n=n_\uparrow+n_\downarrow$. The quantity $\bar w_R$ is the
average value (over the WS cell) of the repulsive part of the Ashcroft empty
core\cite{ash66} pseudopotential,

\begin{equation}
w(r)=-\frac{2z}{r}+w_R,\;\;\;\;\;w_R=+\frac{2z}{r}\theta(r_c-r),
\label{eq11}
\end{equation}
and is given by
$\bar
w_R=3r_c^2/r_s^3$ where, $z$ is the valence of the atom,
$\theta(x)$ is the ordinary step function which assumes the value of unity for
positive arguments, and zero for negative values.

The core radius is fixed to the bulk value, $r_c^B$,
 by setting the pressure of the unpolarized bulk system equal to zero
at the observed equilibrium density
$\bar{n}=3/4\pi[\bar{r}_s^B(0)]^3$:

\begin{equation}
\left.\frac{\partial}{\partial r_s}\varepsilon(r_s,0,r_c)\right|_{
 r_s=\bar r_s^B(0),r_c=r_c^B}=0.
\label{eq12}
\end{equation}
Here, $\bar{r}_s^B(0)\equiv\bar{r}_s^B(\zeta=0)$
is the observed equilibrium density parameter for the unpolarized bulk system,
and takes the value of 3.02 for Ag.
The derivative is taken at fixed $r_c$, and the solution of the above
equation gives $r_c^B$ as a function of
$\bar{r}_s^B(0)$

\begin{equation}
r_c^B[\bar r_s^B(0)]=\frac{1}{3}[\bar r_s^B(0)]^{3/2}\left\{\left[
-2t_s(r_s,0)-\varepsilon_x(r_s,0)+ r_s
\frac{\partial}{\partial
 r_s}\varepsilon_c(r_s,0)-\varepsilon_M(
 r_s)\right]_{r_s=\bar r_s^B(0)}\right\}^{1/2}.
\label{eq13}
\end{equation}

The SJM energy for a spin-polarized system with boundary surface is given by
\cite{pertran}

\begin{eqnarray}
E_{\rm SJM}\left[\nup,\ndo,n_+\right]&=&
E_{\rm JM}\left[\nup,\ndo,n_+\right]+\left(\varepsilon_M(r_s)+\bar
w_R(r_s,r_c^B)\right)\int d\rr\;n_+(\rr) \nonumber \\
  &&+\langle\delta v\rangle_{\rm WS}(r_s,r_c^B)\int
d\rr\;\Theta(\rr)\left[n(\rr)-n_+( \rr)\right],
\label{eq14}
\end{eqnarray}
where
\begin{eqnarray}
E_{\rm
JM}\left[\nup,\ndo,n_+\right]&=&T_s\left[\nup,\ndo\right]+E_{xc}\left[\nup,\ndo
\right] \nonumber\\ &&+\frac{1}{2}\int
d\rr\;\phi\left([n,n_+];\rr\right)\left[n(\rr)-n_+(\rr)\right]
\label{eq15}
\end{eqnarray}
and
\begin{equation}
\phi\left([n,n_+];\rr\right)=2\int
d\rrp\;\frac{\left[n(\rrp)-n_+(\rrp)\right]}{\left|\rr-\rrp\right|}.
\label{eq16}
\end{equation}
Here, $n=n_\uparrow+n_\downarrow$ and $n_+$ is the jellium
density. $\Theta(\rr)$ takes the value of unity
inside the jellium background and zero, outside. The first and second terms in
the right hand side of Eq.(\ref{eq15}) are
the non-interacting kinetic energy and the exchange-correlation energy, and the
last term is the Coulomb interaction energy of the system.
The quantity $\langle\delta v\rangle_{\rm WS}$
is the average of the difference potential over the
Wigner-Seitz cell and the difference potential, $\delta v$, is defined as the
difference between the pseudopotential of a lattice of ions and the
electrostatic potential of the jellium positive background.
The effective potential, used in the self-consistent KS equations, is obtained
by
taking the variational derivative of the SJM energy functional with respect to
the spin densities as

\begin{eqnarray}
v_{eff}^\sigma\left(\left[n_\uparrow,n_\downarrow,n_+\right];\rr\right)&=&
\frac{\delta} {\delta n_\sigma(\rr)}(E_{\rm SJM} -T_s)\nonumber\\
&=&\phi\left(\left[n,n_+\right];\rr\right)+
v_{xc}^\sigma\left(\left[n_\uparrow,n_\downarrow\right];\rr\right)
 +\Theta(\rr)\langle\delta v\rangle_{\rm WS} (r_s,r_c^B),
\label{eq17}
\end{eqnarray}
where $\sigma=\uparrow,\downarrow$.
By solving the KS equations
\begin{equation}
\left(\nabla^2+v_{eff}^\sigma(\rr)\right)\phi_i^\sigma(
\rr)=\varepsilon_i^\sigma
 \phi_i^\sigma(\rr),\;\;\;\;\;\;\;\sigma=\uparrow,\downarrow,
\label{eq18}
\end{equation}

\begin{equation}
n(\rr)=\sum_{\sigma=\uparrow,\downarrow}n_\sigma(\rr),
\label{eq19}
\end{equation}

\begin{equation}
n_\sigma(\rr)=\sum_{i(occ)}\left|\phi_i^\sigma(\rr)\right|^2,
\label{eq20}
\end{equation}
and finding the self-consistent values for $\varepsilon_i^\sigma$ and
$\phi_i^\sigma$, one obtains the total energy.

In our spherical JM, we have

\begin{equation}
n_+(\rr)=\frac{3}{4\pi r_s^3}\theta(R-r)
\label{eq21}
\end{equation}
in which $R=(zN)^{1/3}r_s$ is the radius of the jellium sphere, and $n(\rr)$
denotes the electron density at point $\rr$ in space.
Using the Eq. (21) of Ref. [\ref{pertran}], this average value is given by

\begin{equation}
\langle\delta v\rangle_{\rm WS}(r_s,r_c^B)=\frac{3(r_c^B)^2}{r_s^3}-
\frac{3}{5r_s}.
\label{eq22}
\end{equation}

Applying Eq. (\ref{eq14}) to a metal cluster which contains $N_\uparrow$
spin-up,
 $N_\downarrow$ spin-down and $N$ $(=N_\uparrow+N_\downarrow)$ total electrons
in the ground state, the
 SJM energy becomes a function of $N$,
 $r_s$, and $r_c^B$. The equilibrium density parameter, $\bar r_s(N)$,
for a cluster in the ground state electronic configuration, is the solution of
the equation

\begin{equation}
\left.\frac{\partial}{\partial r_s}E(N,r_s,r_c^B)\right|_{r_s=\bar
r_s(N)}=0.
\label{eq23}
\end{equation}
Here, the derivative is taken at fixed values of $N$  and
$r_c^B$.
For an $N$-electron
cluster in its ground state electronic configuration, we have solved the KS
equations\cite{KohnSham}
self-consistently for various $r_s$ values and obtained
the equilibrium density
parameter, $\bar r_s(N)$, and its corresponding energy,
$\bar E(N)\equiv E(N,\bar r_s(N),r_c^B)$.
\subsection{Dissociation energy and fission barrier}

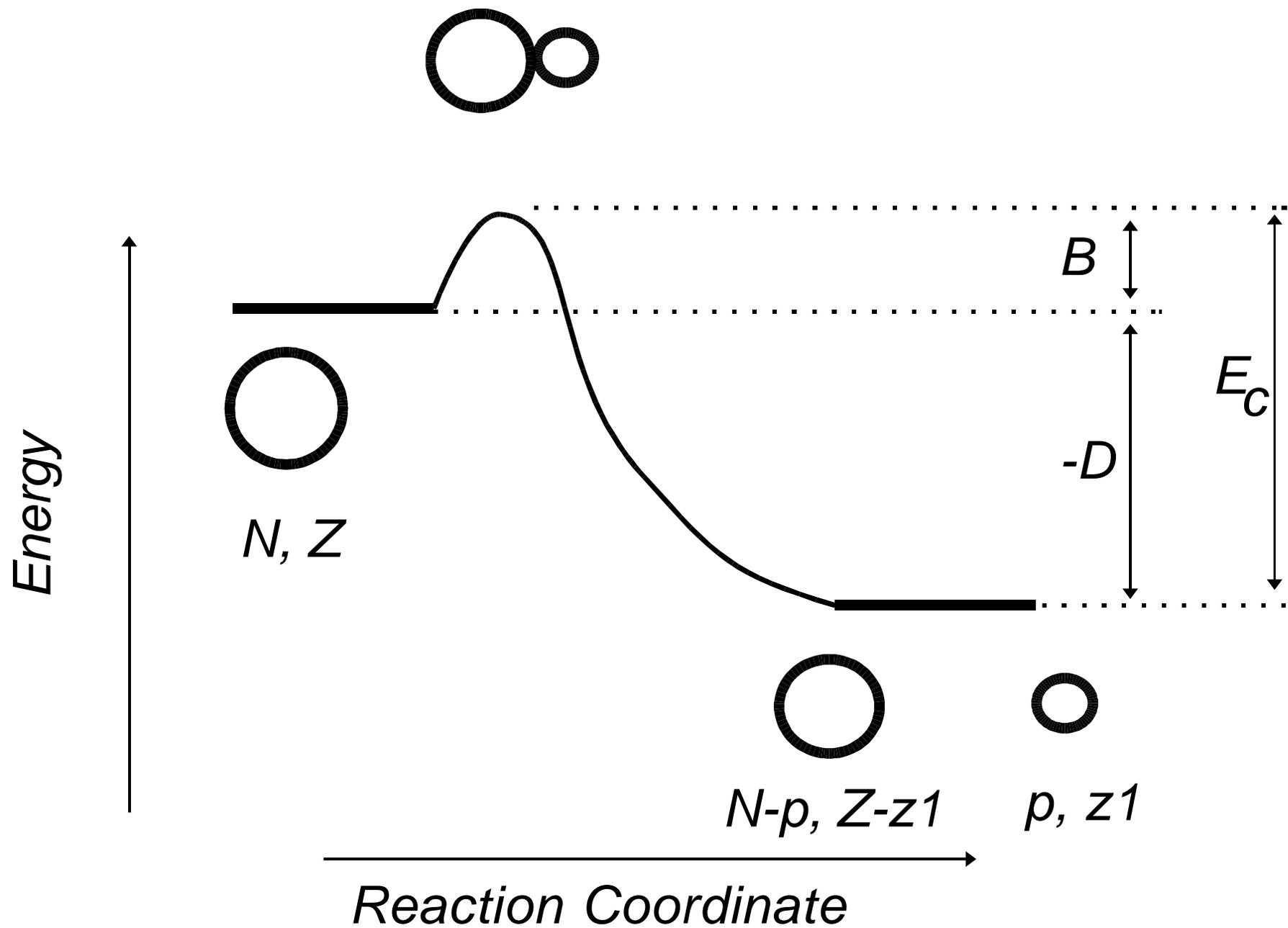
\begin{figure}
\caption{Fission barrier in the two-touching-spheres model. $B$, $D$, and $E_c$
are barrier height, dissociation energy, and the coulomb energy between two
touching charged conductors, respectively.}
\label{fig1}
\end{figure}

The dissociation energy (DE) for the general binary decay process
(\ref{decaygen}), defined as
the difference in the sum of total
energies of the products and the total energy of the parent cluster, is given
by
\begin{equation}
D^Z_{z_1}(N,p)=(E_{N-p}^{Z-z_1} + E_p^{z_1}) - E_N^{Z},
\label{deltagen}
\end{equation}
In evaporation processes ($z_1=0$), a negative value for the DE implies that
the parent
cluster is unstable against that particular decay channel and therefore, the
fragmentation is spontaneous. On the other hand, a positive DE in a particular
decay channel means that the parent cluster is stable against the decay in that
particular channel. That is, one should somehow supply energy to the system to
induce the fragmentation.
For processes (\ref{eq1}) and (\ref{eq2}) the DE becomes
\begin{equation}
D^{1+}_0(N,p)=(E_{N-p}^{1+} + E_p^0) - E_N^{1+}
\label{eq4}
\end{equation}
and
\begin{equation}
D^{2+}_0(N,p)=(E_{N-p}^{2+} + E_p^0) - E_N^{2+},
\label{eq5}
\end{equation}
respectively.
However, in the fission processes ($z_1>0$) as in Eq.(\ref{eq3}), a negative DE
 does not mean that the cluster would decay. It is because of
the existence of a fission barrier which originates from the short-range
attractive (due to binding energy) and long-range repulsive (due to coulomb
repulsion) forces between the charged products. The situation is shown in
Fig. \ref{fig1}. Any excitation above the
barrier, which may be induced by collisions or radiation, will eventually make
the expected decay possible. One of the main deficiencies of the JM is that it
gives negative values\cite{lang70} of surface energies for $r_s\le 2$. Our
using of SJM-SC
is expected, therefore, to give more realistic values of surface energies and
barrier heights.
The fission barrier $B_{z_1}^Z(N,p)$ is approximated by the Coulomb interaction
$E_c$ of two touching spheres (i.e., the fission products) and the DE as

\begin{equation}
B_{z_1}^Z(N,p)=D^Z_{z_1}(N,p)+E_c
\label{eq24}
\end{equation}

For the Coulomb interaction between the two fission products we take into
account their polarizabilities. The interaction energy of two charged
conducting spheres can be calculated numerically using image charge
method.\cite{Naher} An equally good but much simpler approach is the use of the
analytical expression\cite{Bott,Kruck} for the interaction between charges
$z_1$
and $z_2$ with polarizabilities $\alpha_1$ and $\alpha_2$ at a distance $s$

\begin{equation}
E_{1,2}^{\rm
pol}(s)=\frac{A_2}{A_4}\frac{2z_1z_2}{s}
-\frac{\alpha_1}{s^3}\frac{1}{A_4}\frac{z_2^2}{s}
-\frac{\alpha_2}{s^3}\frac{1}{A_4}\frac{z_1^2}{s}.
\label{eq25}
\end{equation}
Here $A_j$ is given by

\begin{equation}
A_j=1-\frac{j\alpha_1\alpha_2}{s^6},
\label{eq26}
\end{equation}
and the polarizability of a conducting sphere (i.e., the metal cluster) with
radius $R$ is \cite{Uwe} $\alpha=R^3$. An other formula which was
used \cite{Naher,Nakamura} for the Coulomb interaction of two touching
conducting spheres is given by

\begin{equation}
E_{1,2}=\frac{2z_1z_2}{R_1+R_2+2\delta R}
\label{eq27}
\end{equation}
where, for silver, the value $\delta R=0.94 $ takes the polarizability into
account. The BH's for small clusters, obtained from this formula,
are
somewhat smaller than those we obtained using Eqs. (\ref{eq25}), (\ref{eq26}).
Koizumi {\it et. al.}\cite{Koizumi1,Koizumi2} have calculated the barrier
heights for the fission of doubly charged silver clusters using a shape
function in the LDM with shell correction.

\section{Results and discussion}
\label{sec3}

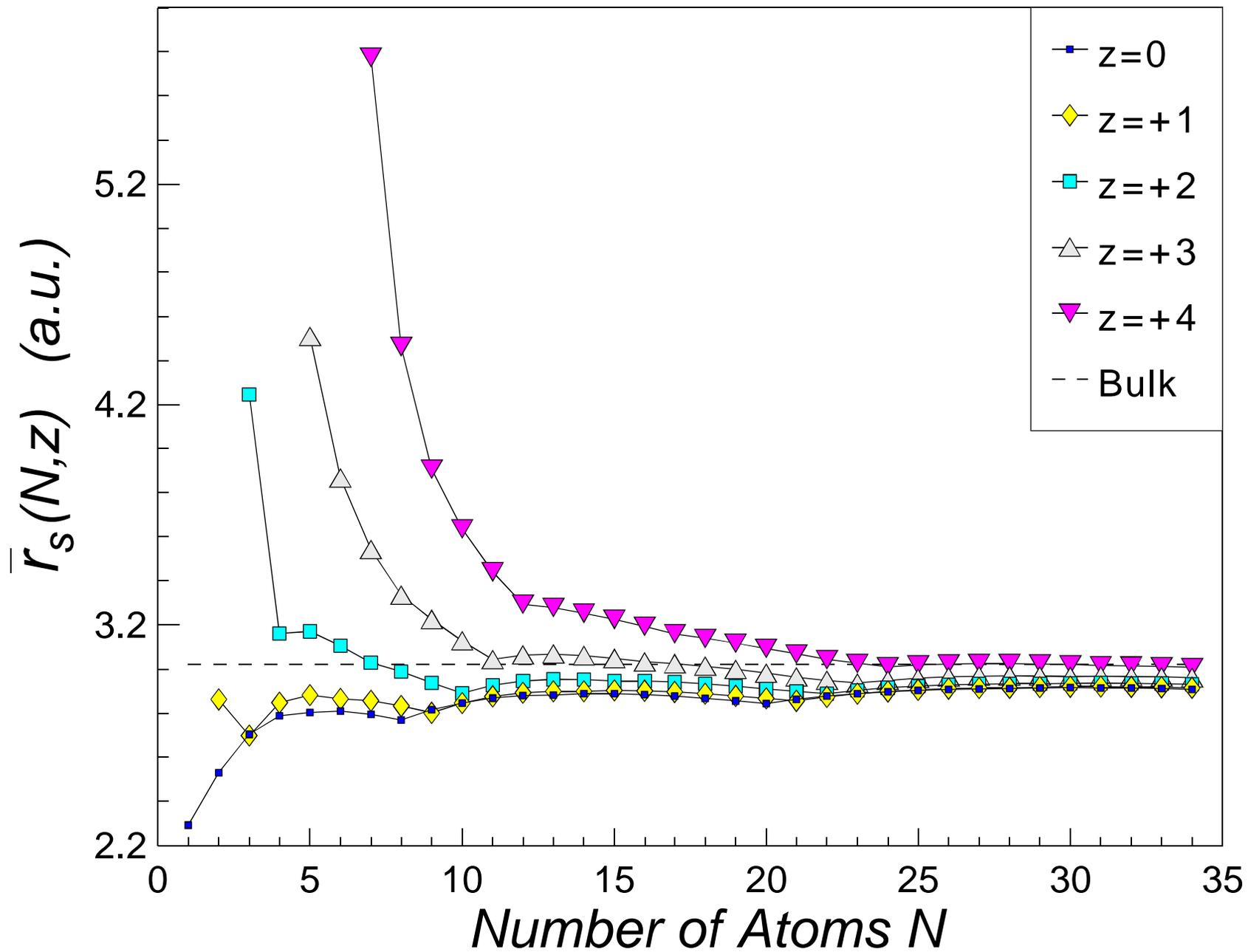
\begin{figure}
\caption{The equilibrium $r_s$ values in atomic units for $Z$-charged clusters
as a function of the cluster size $N$. The dashed line is the bulk Wigner-Seitz
radius for Ag ($r_s=3.02$).}
\label{fig2}
\end{figure}

After an extensive self-consistent SJM-SC calculations, we have calculated
the equilibrium $r_s$ values and the energies of Ag$_N^Z$ ($Z$=0,1,2,3,4)
 for different
cluster sizes
$(1\le N\le 100)$. To show the main differences in the equilibrium $r_s$ values
of these clusters, which are appreciable for
relatively small clusters, we have plotted, in Fig. \ref{fig2}, the
corresponding
$\bar r_s(N)$ values only up to $N=34$. As is obviously seen in the figure, the
neutral and singly ionized clusters are self-compressed for all values of $N$.
This is because of the dominant effect of the surface tension.
 However, for multiply charged clusters, the $\bar r_s(N)$ values
cross the bulk border (i.e., $r_s=3.02$) at some $N$ which we show it by $N_0$.
Our results show that, in general, for
larger values of charging, the self-expansion persists up to larger values of
$N_0$. That is, $N_0^{2+}<N_0^{3+}<N_0^{4+}<\cdots$. For example, here we have
obtained the values of 7, 17, 23 for $N_0^{2+},\;N_0^{3+},\;N_0^{4+}$,
respectively. This means that, for larger charging values, the coulomb
repulsion
between excess charges dominates the surface tension up to larger values
of $N_0$.
It is also clearly seen in Fig. \ref{fig2} that for clusters with the
same numbers of electrons but different numbers of atoms, $N$, the
following inequality holds

\begin{equation}
\bar r_s^0(N)<\bar r_s^{1+}(N+1)<\bar r_s^{2+}(N+2)<\bar
r_s^{3+}(N+3)<\cdots.
\label{eq28}
\end{equation}
For Ag$_N^{4+}$ clusters, we could not find any solution
of the Eq. (\ref{eq23}) for $N=5$. That is, the single remaining electron is
not able to bind the 5 constituent ions to each other in the Ag$_5^{4+}$
system. However,
the solutions of $\bar r_s^{4+}(6)=14.76$ and $\bar r_s^{3+}(4)=29.0$ have been
obtained for Eq. (\ref{eq23}) which are so large that one can not realize the
corresponding bound states experimentally and we ignore these bound states.

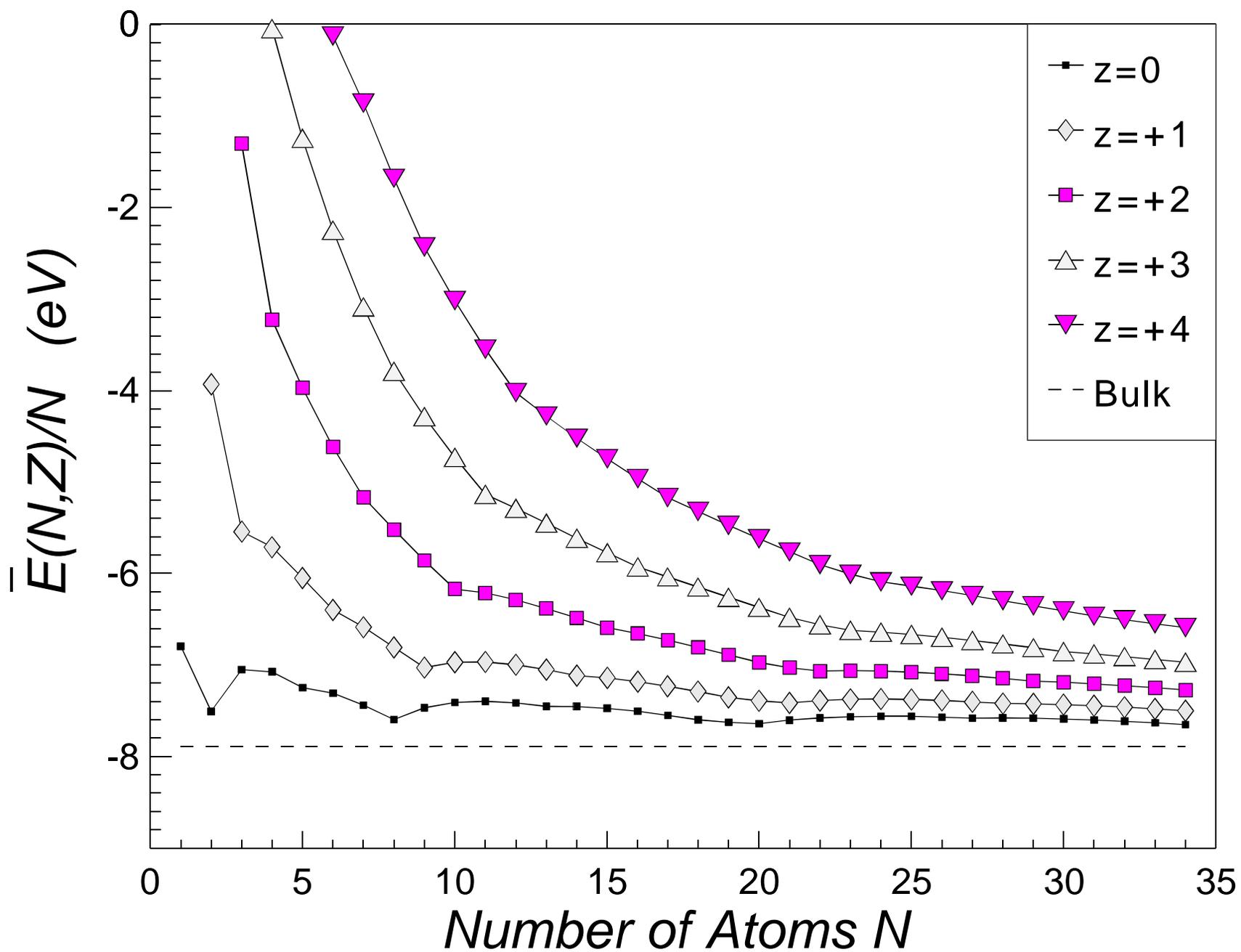
\begin{figure}
\caption{The equilibrium total energies per atom in electron volts for
$Z$-charged
clusters as a function of the cluster size $N$. The dashed line is the bulk SJM
value for Ag ($\varepsilon=-7.89 eV$).}
\label{fig3}
\end{figure}

Figure \ref{fig3} shows the equilibrium energies per atom in electron-volts
for Ag$_N^0$, Ag$_N^{1+}$, Ag$_N^{2+}$, Ag$_N^{3+}$, and
Ag$_N^{4+}$ with different cluster sizes ($1\le N\le 34$). For comparison, we
have also plotted the bulk value ($\varepsilon=-7.89eV$) by a dashed line.
As is seen, by increasing the charge of a given $N$-atom cluster, the
coulomb repulsion between the excess charge induces an inflation in the
cluster (see Fig. \ref{fig2}) and therefore, the density of the material in the
cluster decreases which, in turn, leads to a smaller binding energies.

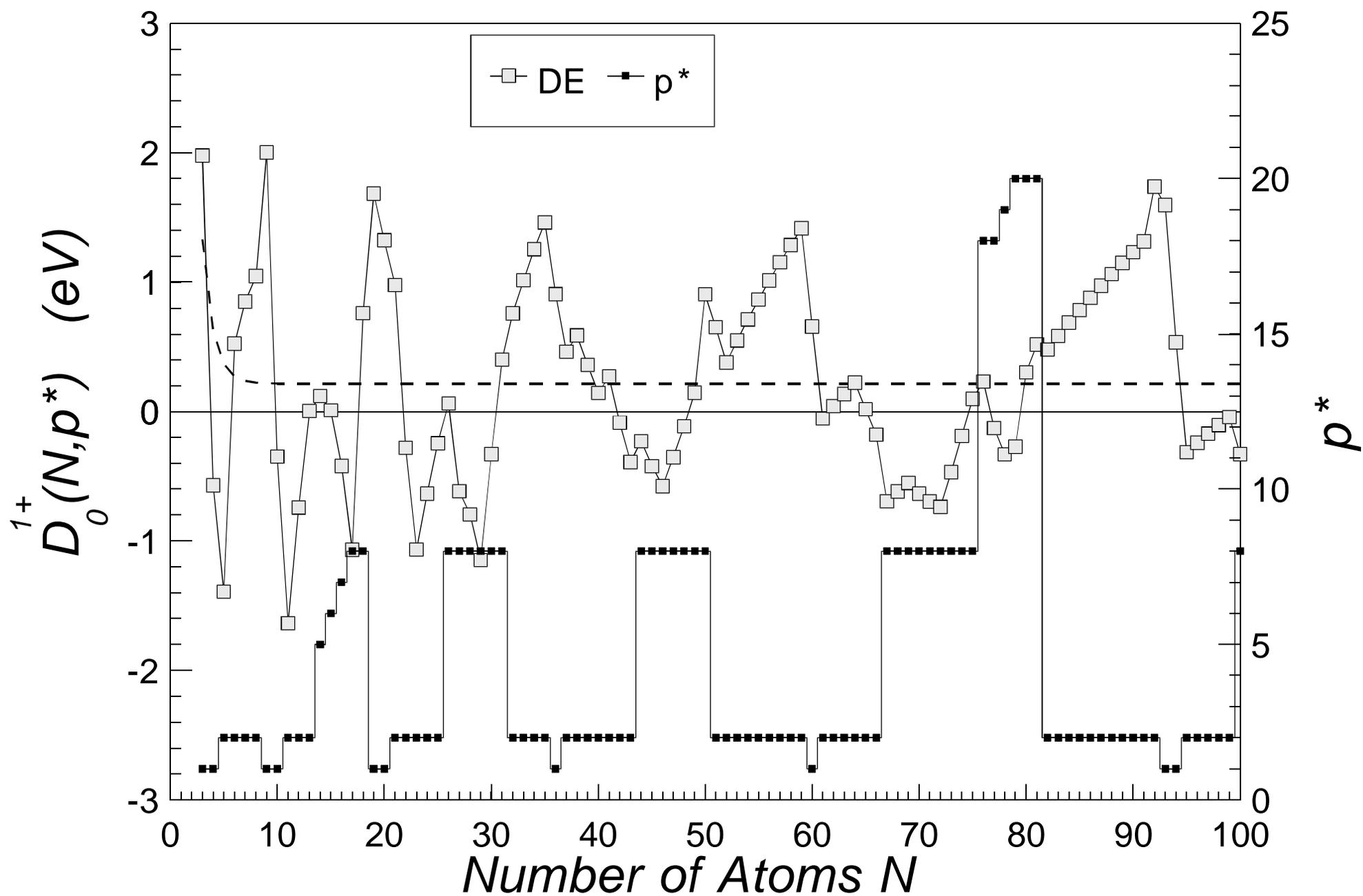
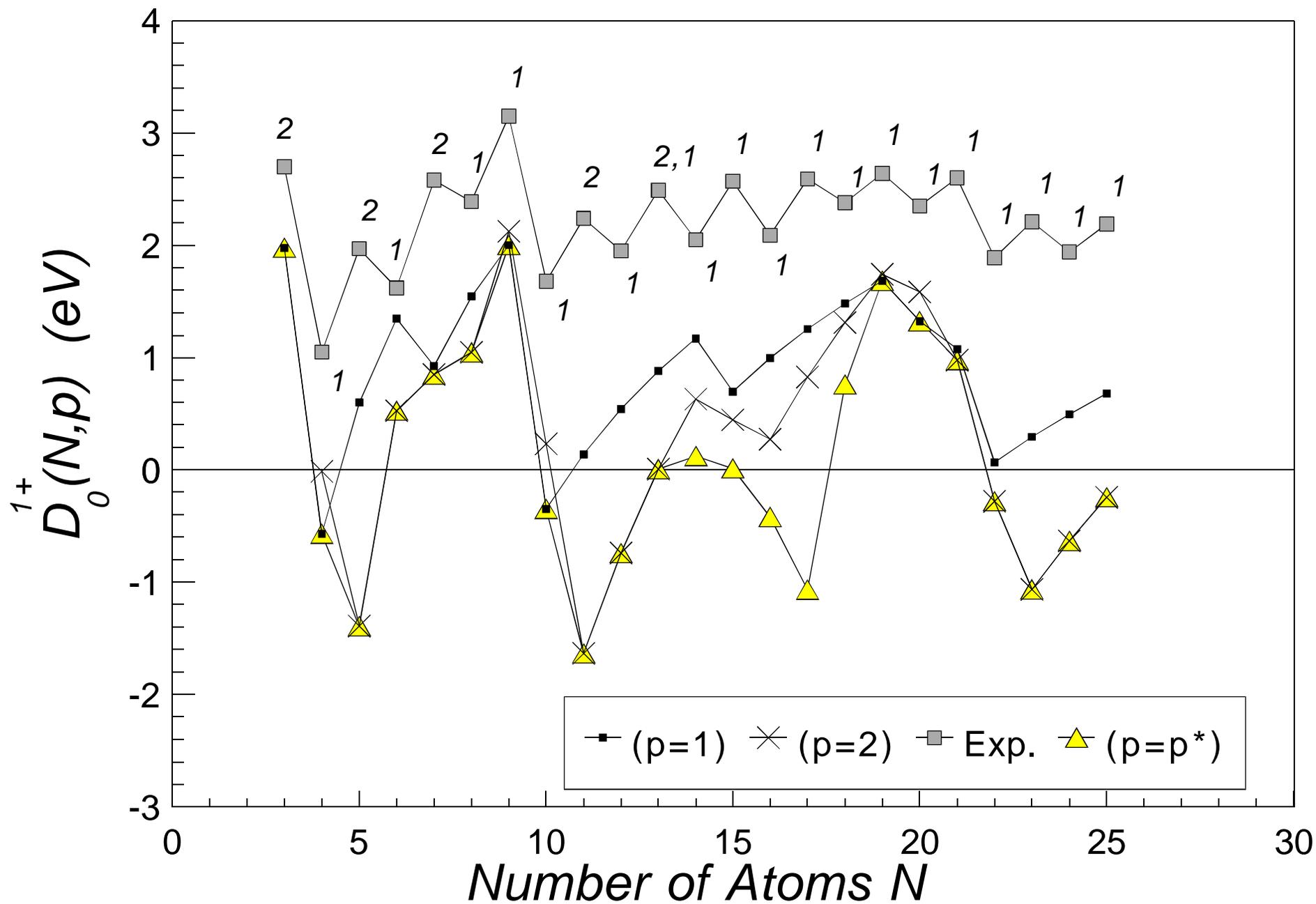
\begin{figure}
\caption{(a)- The right vertical axis shows the size of the fragment in the
most favored channel, $p^*$, of singly ionized clusters by small solid squares.
The left vertical axis shows
the dissociation energies, DE, in electron volts for the most favored decay
channels of singly ionized clusters by large open square symbols.
The dashed line is the fitting result which
has the asymptotic value of $0.20eV$.
(b)- The dissociation energies for monomer ($p=1$), dimer ($p=2$), and the most
favored ($p=p^*$) $p^*$-mers are compared with the experiment. The figures on
the
experimental data points specify the fragment size $p$. The experiment shows an
odd-even staggering.}
\label{fig4}
\end{figure}

In Fig. \ref{fig4}(a) we have plotted the DE's of the most
favored decay channels for the process
${\rm Ag}_N^{1+}\to {\rm Ag}_{N-p}^{1+} + {\rm Ag}_p^0$. By definition, the DE
is minimum in the most favored channel. We have shown the most favored
value of $p$ by $p^*$. The solid small square symbols show the most favored
values $p^*$ on the right vertical axis whereas, the corresponding DE's,
$D^{1+}_0(N,p^*)$, are shown on the left vertical axis by large open squares.
The dashed line is the result of a fitting to the quantal
DE's. As is seen, the magority of the clusters have positive
DE's and therefore, they are stable against the spontaneous
decay. However, the remaining clusters have negative DE's and accorgingly, they
decay into smaller fragments. Clusters close to the closed-shell ones, decay by
emitting a monomer or dimer. On the other hand, clusters that are far from
being a closed-shell, can break into two fragments each of which are close
or identical to closed-shell ones. For example, {\rm Ag}$_3^{1+}$ emits a
neutral monomer and the remaining is a singly charged dimer, {\rm
Ag}$_{44}^{1+}$
emits an {\rm Ag}$_8^0$ which is a closed-shell and the remaining is
{\rm Ag}$_{36}^{1+}$ which is close to a closed-shell, and finally,
{\rm Ag}$_{80}^{1+}$ emits {\rm Ag}$_{20}^0$ and the situation is similar to
the latter one. Except for the closed-shell singly ionized cluster
{\rm Ag}$_{69}^{1+}$, all other closed-shell singly ionized clusters,
{\rm Ag}$_N^{1+}$ ($N$=3, 9, 19, 21, 35, 41, 59, 91, 93) are stable against the
spontaneous decay. The dashed fitted line which resembles the result
of liquid-drop model (LDM) calculations ( see Fig. 4 of Ref. [\ref{Vieira95}]),
predicts that all singly ionized clusters are stable and the
asymptotic value of DE is constant and equal to $0.20eV$.

Figure \ref{fig4}(b) compares the experimental\cite{Kruck99} dissociation
energies with the monomer DE's $D^{1+}_0(N,1)$, dimer DE's $D^{1+}_0(N,2)$,
and the most favored DE's $D^{1+}_0(N,p^*)$. As is seen, the most favored
fragments are somewhere monomers, somewhere dimers and somewhere none of them.
The general trend of the calculated monomer dissociation energies is similar
to the experimental one and has a better agreement with the experiment than the
other two DE's. That is, from $N=3$ to $N=4$ the energy decreases; from $N=4$
to $N=9$ the energy increases in the mean; a decrease on going from 9 to 10; an
increase from 10 to 21; and finally, a decrease from $N=21$ to $N=22$ and again
increasing from $N=22$. However, our results lack the odd-even staggering
because, it originates from the nonspherical shapes for the
jellium. Resorting to non-spherical shapes also decreases the
pronounced shell effects.\cite{Nakamura}
The relative smallness of our calculated DE's can be explained in terms of the
details of the experimental setup. If the experiment starts using  neutral
Ag$_N$ clusters, then the equilibrium $r_s$ values would be smaller than the
bulk value (see the $Z=0$ plot in Fig. \ref{fig2}), and therefore, the total
energies would be more negative (see the $Z=0$ plot in Fig. \ref{fig3}).
Now, irradiating the parent neutral cluster with a high power laser beam would
lead to the ionization of the neutral cluster. If the
photons also interact with the ionized cluster before the ionized cluster
achieves its relaxed state, then the equilibrium $r_s$ value of the
ionized cluster would be less than the relaxed value (In our calculations
we have used the relaxed values at all steps.), and therefore, the magnitude of
the total energy of the ionized parent cluster would be larger. This fact would
lead to larger values of the DE's. Comparing the experimental data with our
results we conclude
that the photo-dissociation occurs before the relaxation of the parent ionized
cluster is completed. The other extreme is that we consider the `{\it sudden}'
approximation in which we assume that the relaxation time for the ionized
cluster is infinite and the cluster undergoes the dissociation without changing
the volume (i.e., the saturation approximation which is used in nuclear
fragmentations and ordinary jellium calculations for clusters). In reality,
neither of these extreme `{\it relaxed}' or `{\it sudden}' approximations are
at work but something in between.

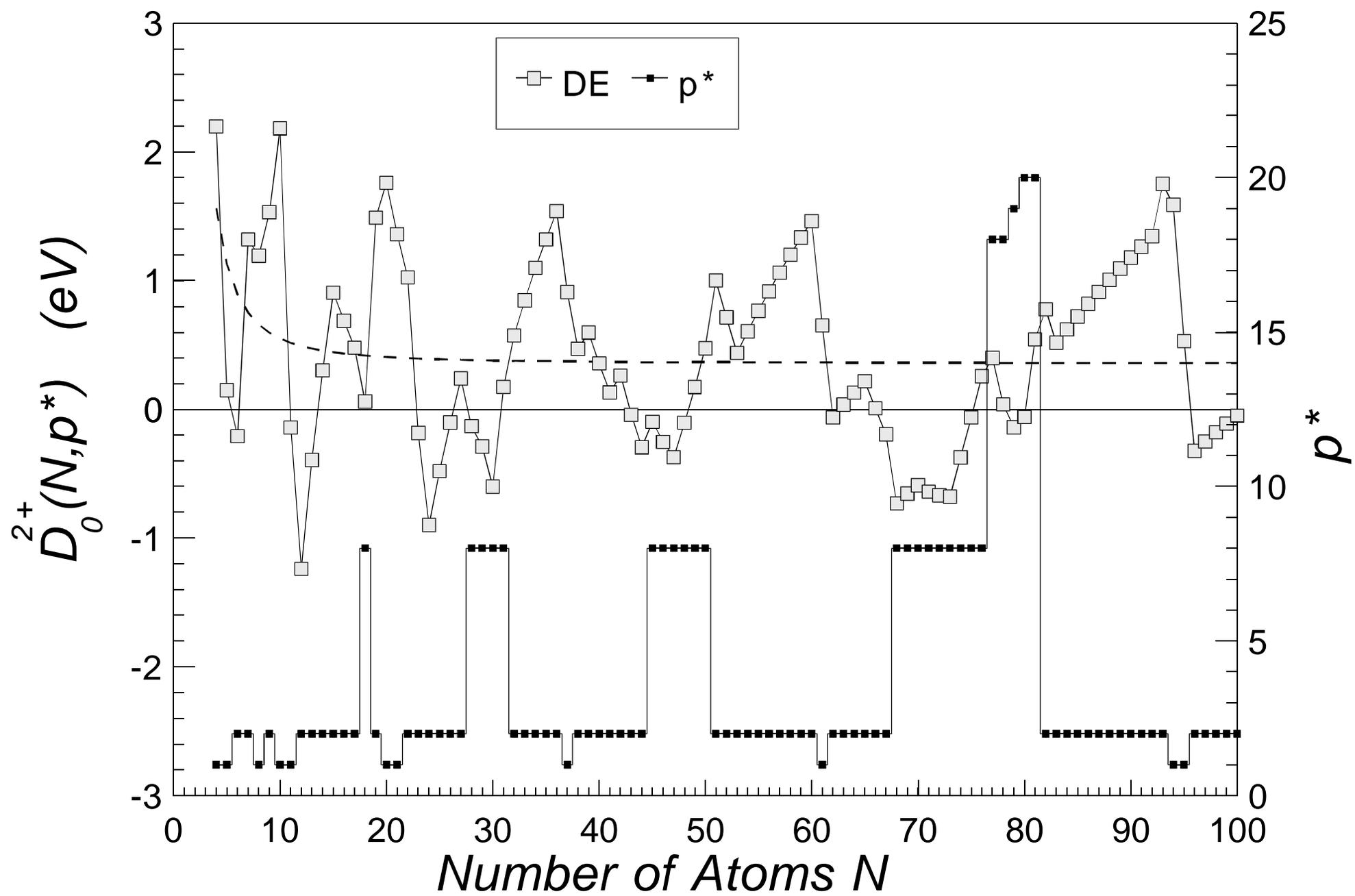
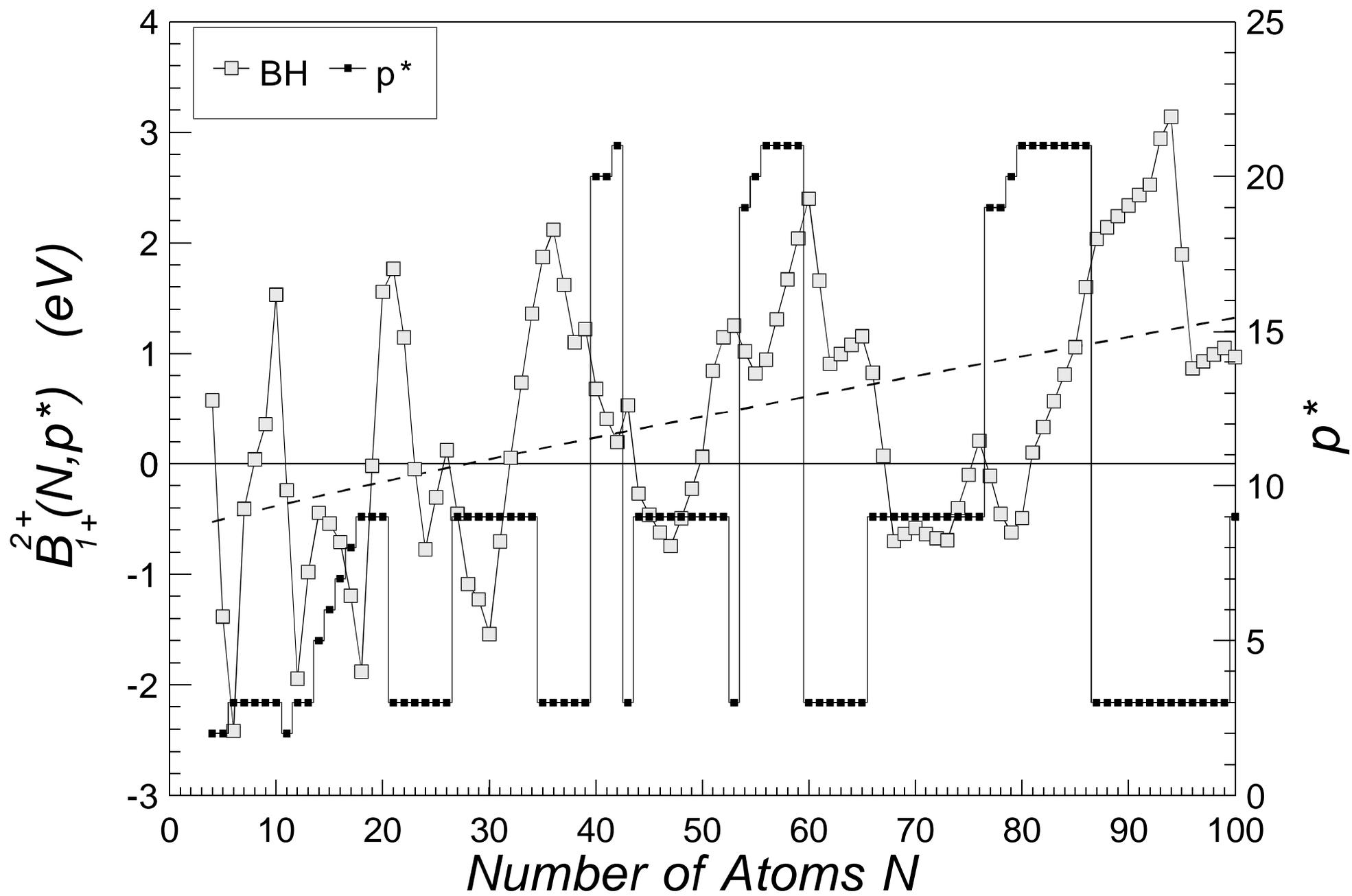
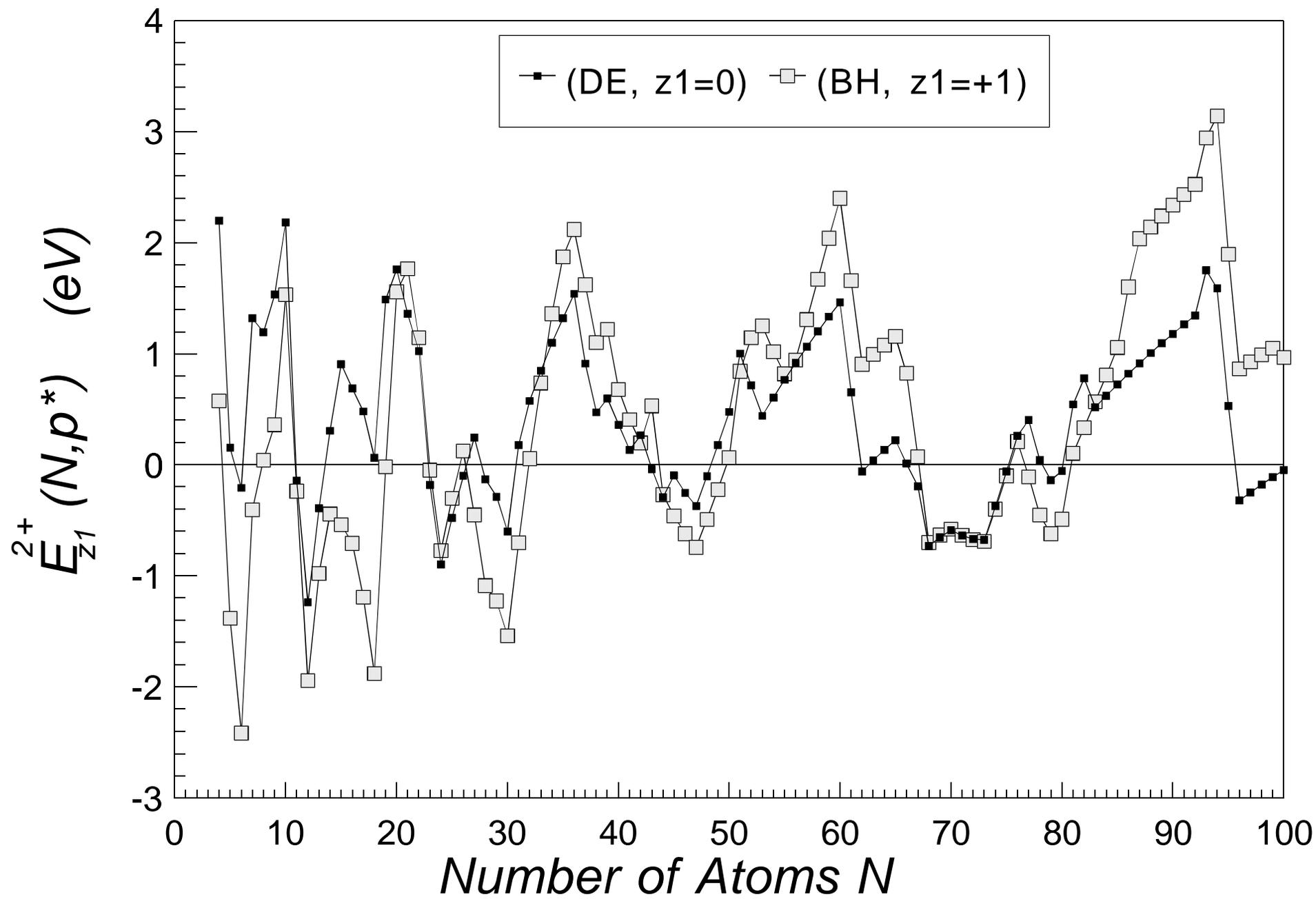
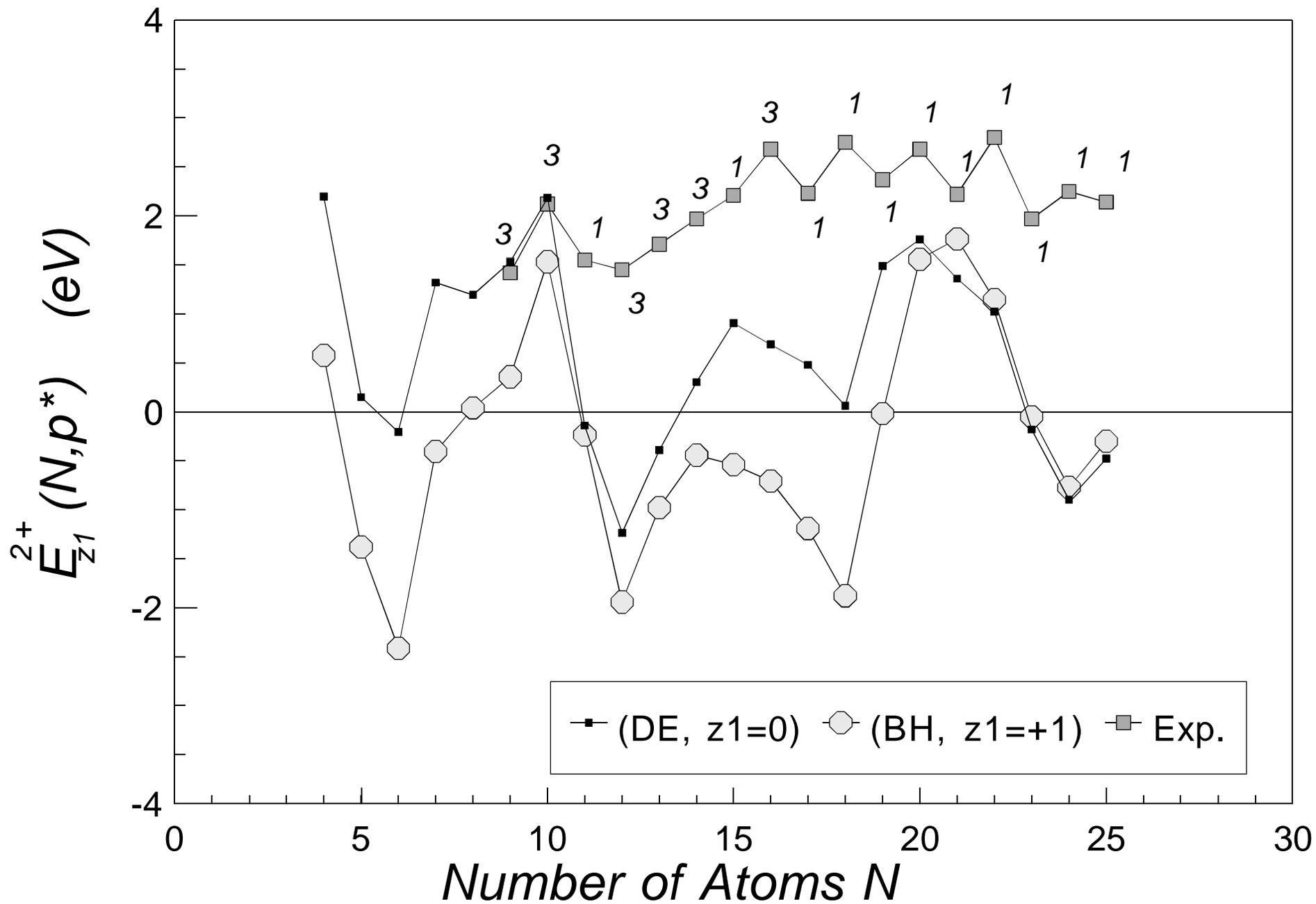
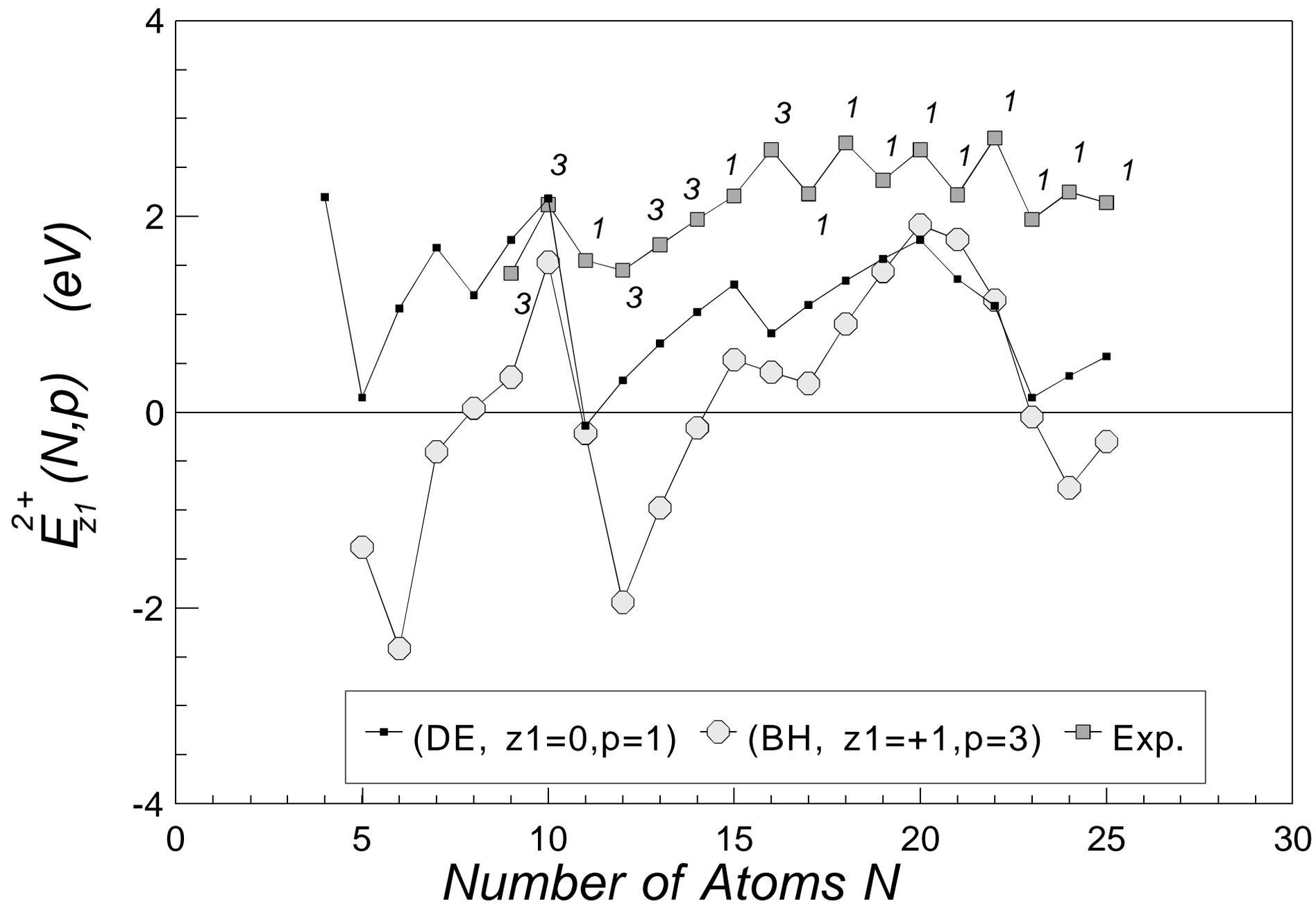
\begin{figure}
\caption{(a)- The same as in Fig. \protect\ref{fig4}(a) for doubly ionized
clusters, with the asymptotic value of the dashed fitted line as $0.40 eV$.
(b)- The most favored size of the fragments, $p^*$, and the barrier heights for
this most favored values in electron volts as function of the cluster size $N$.
The dashed fitted line intersects the zero axis at $N=29$.
(c)- Comparison of the most favored values of the DE's and the BH's.
$E^{2+}_{z_1}$ is $D^{2+}_0$
for $z_1=0$ and $B^{2+}_{1+}$ for $z_1=1$. Evaporation starts from $N=21$.
(d)- Comparison of the most favored DE's and BH's with the experiment. The
figures on the experimental data points show the fragment sizes.
(e)- Comparison of the monomer DE's and the singly ionized trimer BH's with the
experiment.}
\label{fig5}
\end{figure}

In Fig. \ref{fig5}(a) we have shown the most favored products Ag$_{p^*}^0$
and the dissociation energies $D^{2+}_0(N,p^*)$ for the decay of
{\rm Ag}$_N^{2+}$ via evaporation
channel. For this process, as in singly ionized case,
the dashed fitted line predicts no spontaneous decays and shows a
higher constant asymptotic DE as $0.40 eV$. The most favored products are
mainly monomers, dimers and octamers.

Figure \ref{fig5}(b) shows the barrier heights $B^{2+}_{1+}(N,p^*)$ for
 the most  favored channels of the process
${\rm Ag}_N^{2+}\to {\rm Ag}_{N-p}^{1+} + {\rm Ag}_p^{1+}$. By definition,
the most favored fission channel has a minimum value for the BH. As is seen,
some of the BH's are negative. The negativity of
a BH means that we need no energy to supply the system to initiate the fission.
The dashed line which shows the mean behavior of BH intersects the
zero line at $N_a^{\rm mean}\approx 29$ (the mean appearance size). This means
that on the average,
all Ag$_N^{2+}$ clusters with $N<29$ are unstable against spontaneous fission.
The values of $p^*$ show that
most of the emitted fragments are closed-shell {\rm Ag}$_N^{1+}$ clusters with
$N=3,9,21$.

In Fig. \ref{fig5}(c) we have compared the most favored decays of Figs.
\ref{fig5}(a) and \ref{fig5}(b). It is clearly seen that in a certain size
range, the fission and evaporation definitly start their competition.
Our quantal results in
Fig. \ref{fig5}(c) show that in the size range $21\le N\le 26$ the evaporation
dominates the fission which is in good agreement with the Katakuse {\it et.
al.} experimental results\cite{Katakuse} that reveal fission for $N\le 22$.
However, our result is
slightly larger than the Kr\"uckeberg {\it et. al.}\cite{Kruck} experimental
data which show that the fission occurs for $N\le 16$. This difference in the
experimental results depends on the details of the experiment. For $N>26$, our
results show that fission dominates again.
To estimate the size range at which evaporation completely dominates the
fission, we simply find the intersection point of the two mean
behaviors (dashed lines) in Figs. \ref{fig5}(a) and \ref{fig5}(b). A simple
calculation gives this mean critical value as $N_c^{\rm mean}\approx 50$. That
is, in an induced
fragmentation experiment of Ag$_N^{2+}$ clusters, the evaporation dominates the
fission for $N>50$.

In Fig. \ref{fig5}(d) we have compared the most favored values
$D^{2+}_0(N,p^*)$
and $B^{2+}_{1+}(N,p^*)$ with the experimental threshold energies\cite{Kruck}.
Here, also we have smaller DE's and BH's compared to the experiment.
One reason
for this behavior is that the equilibrium volume of the parent cluster is not
equal to the sum of the equilibrium volumes of the product clusters (i.e., the
`{\it relaxed}' approximation) but is
larger. The larger value of the equilibrium $r_s$ leads to a smaller magnitude
of the initial energy and therefore, by Eq. (\ref{eq24}) to a smaller barrier
heights. In other words, the energy needed to deform the parent cluster toward
the fission ( In deformation the surface area increases.) is partly paied as a
 result of self-expansion of the parent cluster.
Our calculations show that in almost all decay channels, the sum of volumes
of the decay products is smaller than that of the parent cluster which
can be explained by the fact that in smaller clusters the surface effect
is higher than that in larger clusters.
The calculated results for the most favored values of DE's at $N=9$ and $N=10$
are very close to the experimental values. The most favored products at $N=9$
and $N=10$ are neutral dimer and monomer, respectively (see Fig.
\ref{fig5}(a)).

In Fig. \ref{fig5}(e) we have compared the calculated monomer DE's
$D^{2+}_0(N,1)$ and the singly charged trimer BH's $B^{2+}_{1+}(N,3)$ with the
experiment. Here also the difference is appreciable.

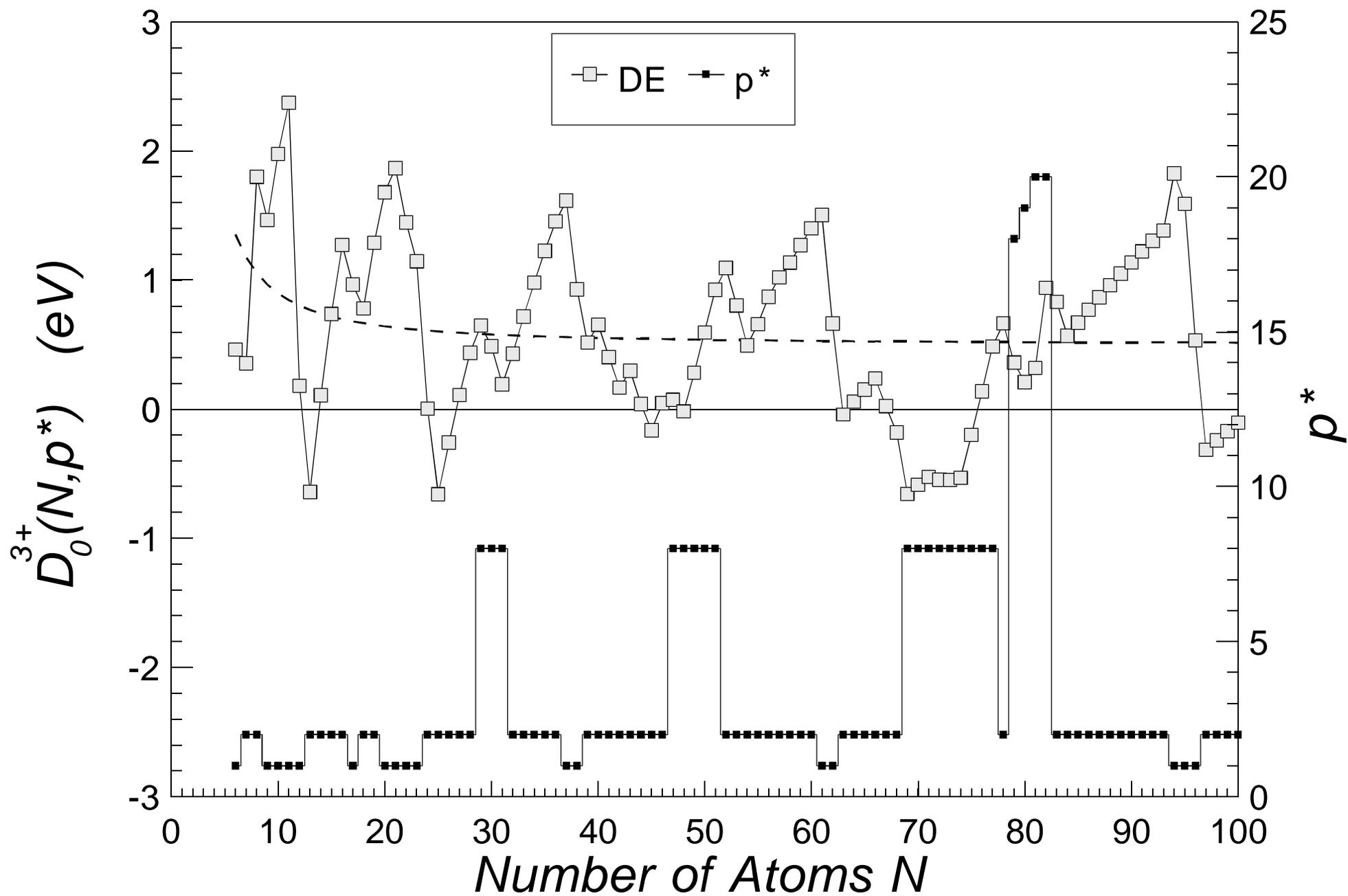
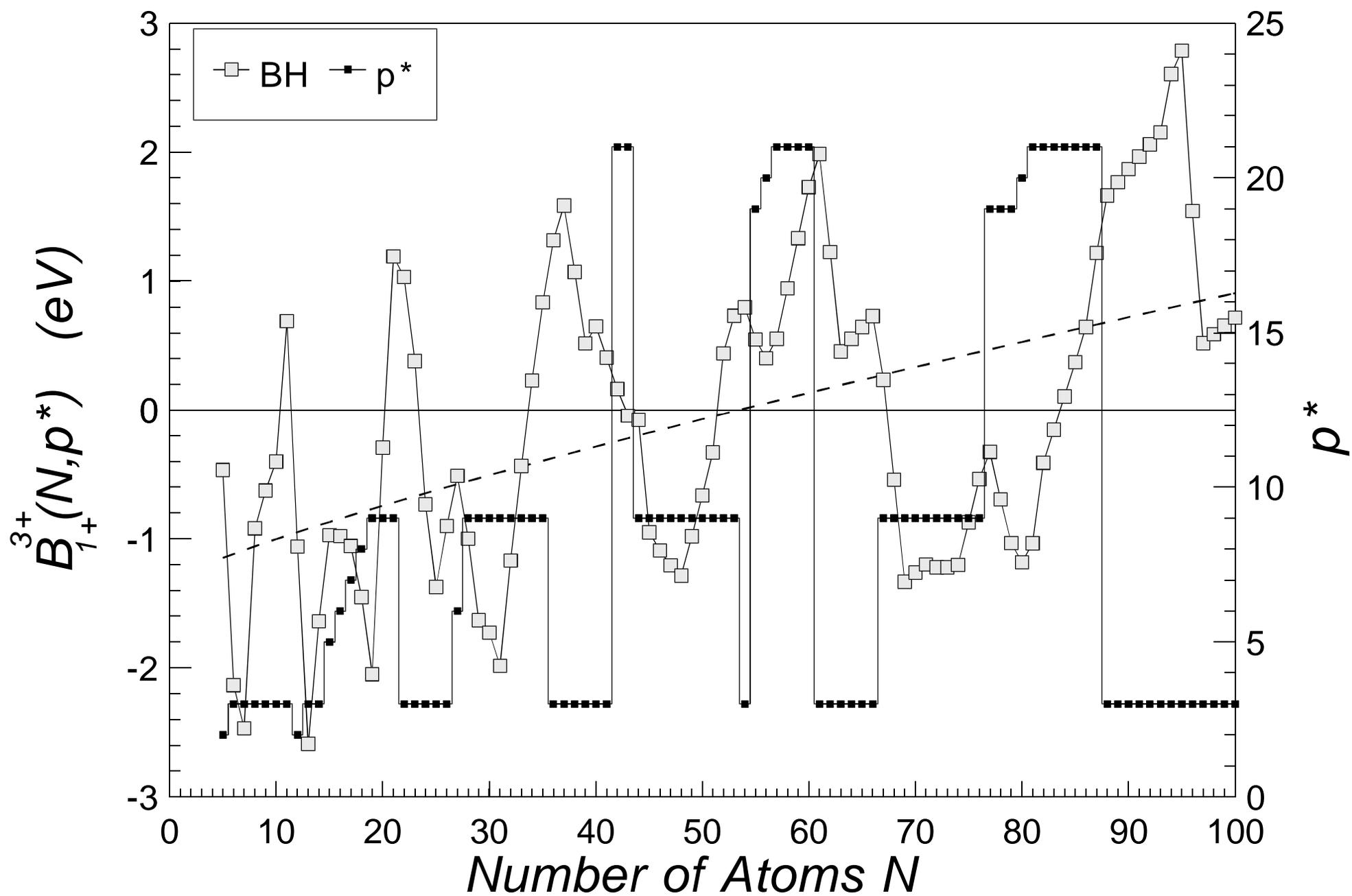
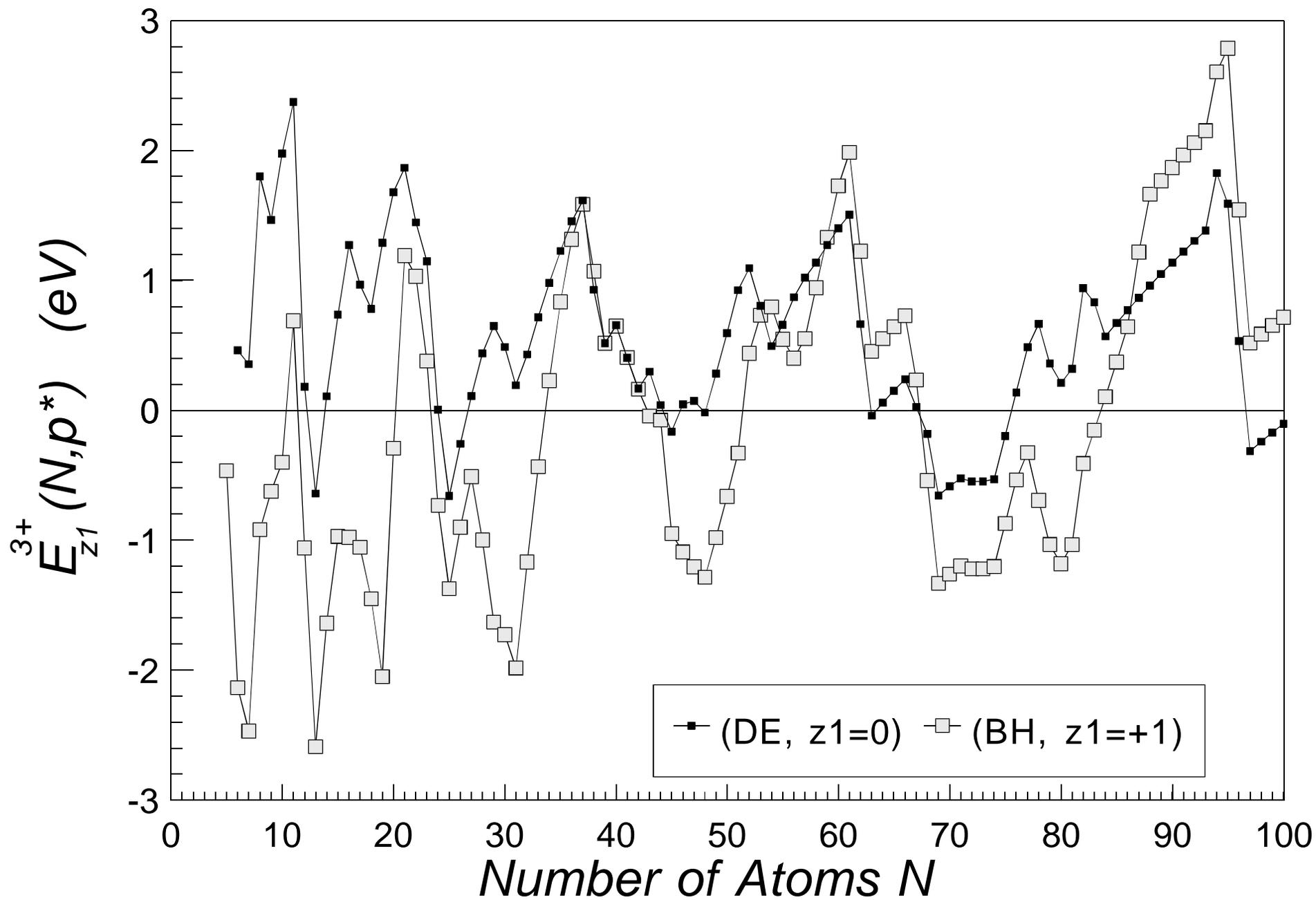
\begin{figure}
\caption{(a)-
The same as in Fig. \protect\ref{fig4}(a) for triply ionized
clusters, with the asymptotic value of the dashed fitted line as $0.50 eV$.
(b)- The same as in Fig. \protect\ref{fig5}(b) with the intersection point at
$N=53$.
(c)- The same as in Fig. \protect\ref{fig5}(c) with starting the evaporation
from $N=38$.}
\label{fig6}
\end{figure}

In Fig. \ref{fig6}(a) we have plotted the $p^*$ and the $D^{3+}_0(N,p^*)$
 for different cluster sizes. The situation is similar
to other previous evaporation processes.
 Here also the asymptotic behavior of the
fitted line predicts no decay and has a constant value of about $0.50 eV$.

In Fig. \ref{fig6}(b), we have plotted the BH's
$B^{3+}_{1+}(N,p^*)$ for the most favored
channel of the binary fission of the process
${\rm Ag}_N^{3+}\to{\rm Ag}_{N-p}^{2+}+{\rm Ag}_p^{1+}$. Here, the fission
products with smaller charge are more or less the same as those in the
 fission of
Ag$_N^{2+}$ clusters. The mean behavior dashed line intersects the zero axis at
$N_a^{\rm mean}\approx 53$. The slope of this line is larger than that of Fig.
\ref{fig5}(b).

Figure \ref{fig6}(c) compares the most favored decays of Figs.
\ref{fig6}(a) and \ref{fig6}(b). To our knowledge, there is no experimental
results in the literature on the decay of Ag$_N^Z$ with $Z\ge 3$. It is seen
that at $N=38$, evaporation dominates and from $N=39$ to $N=42$
evaporation and fission are equally probable. From $N=43$ to $N=58$, except for
$N=54$, fission dominates again. From $N=59$ to $N=67$ the evaporation process
overcomes and so on. To obtain the mean critical value, we find
the intersection point of the two mean
behaviors (dashed lines) in Figs. \ref{fig6}(a) and \ref{fig6}(b) which
results in the value $N_c^{\rm mean}\approx 80$.

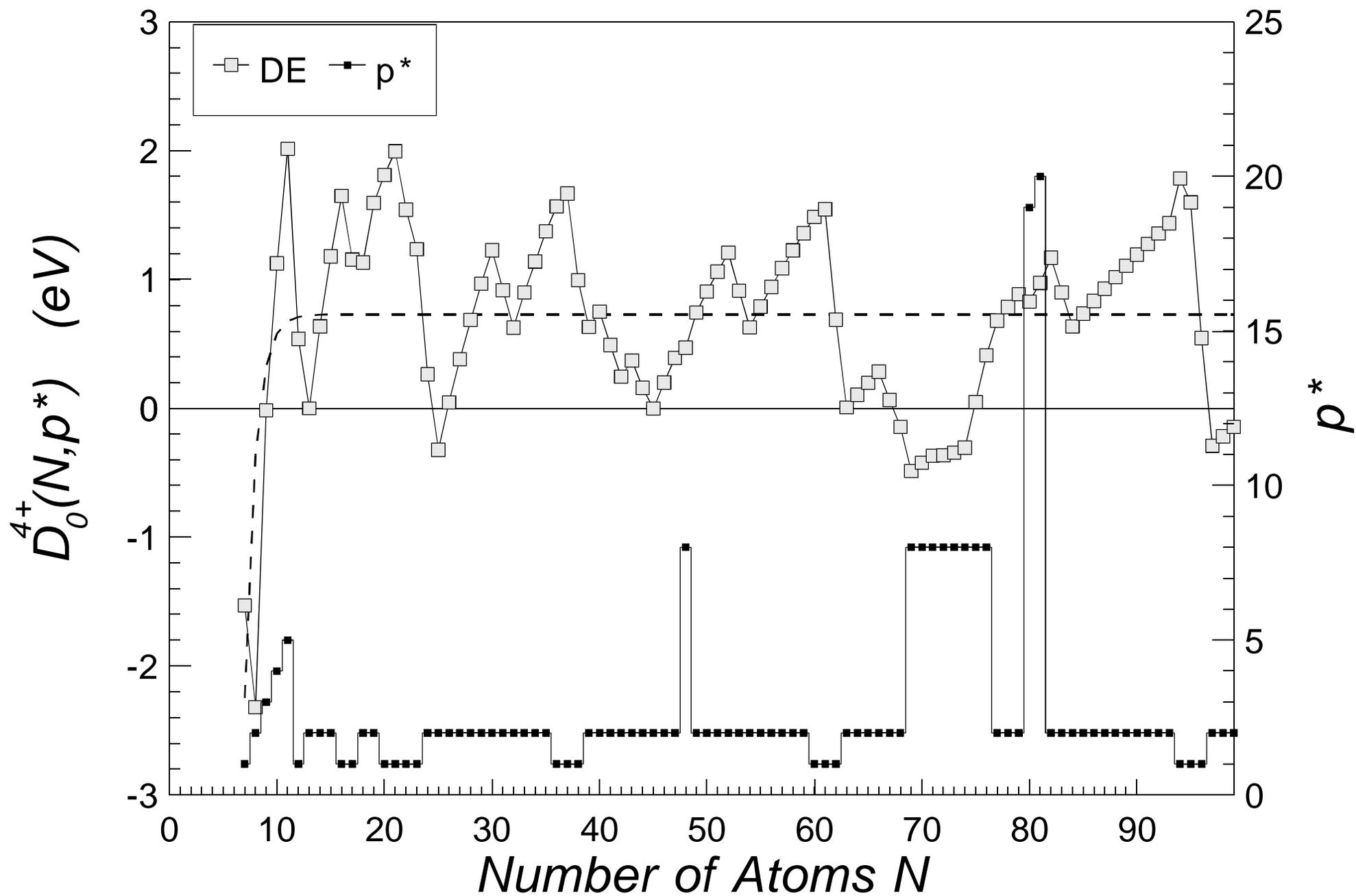
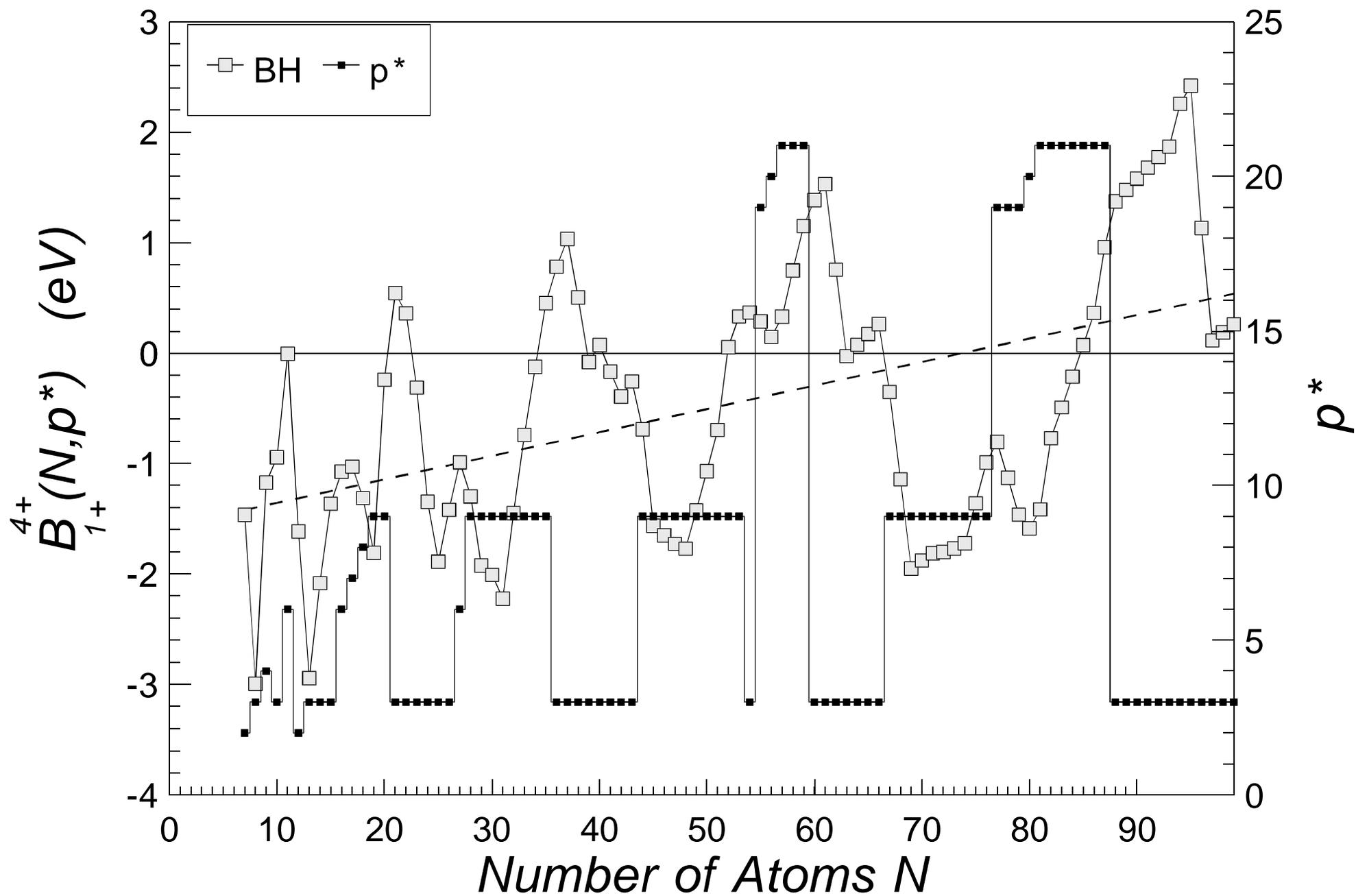
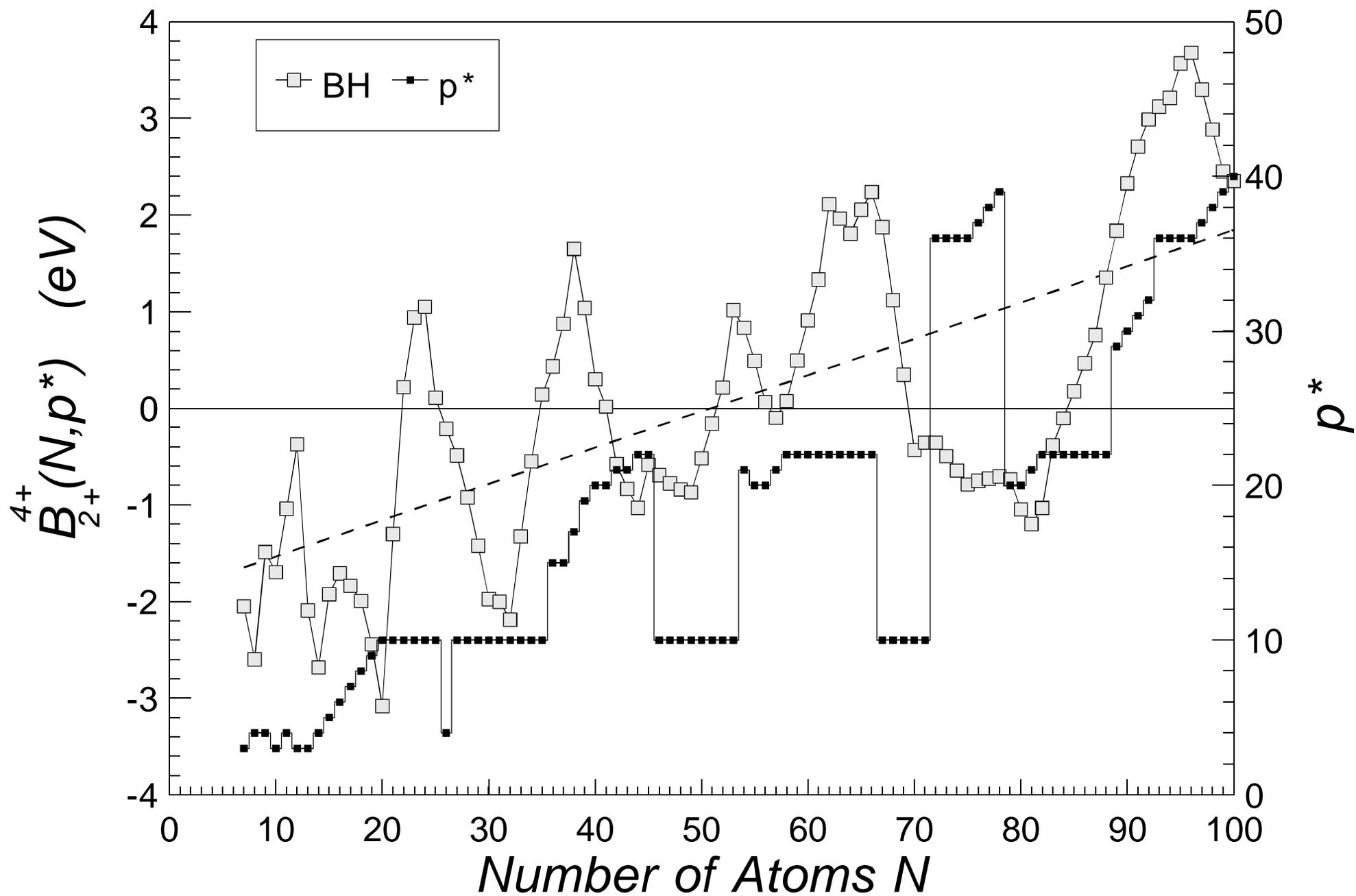
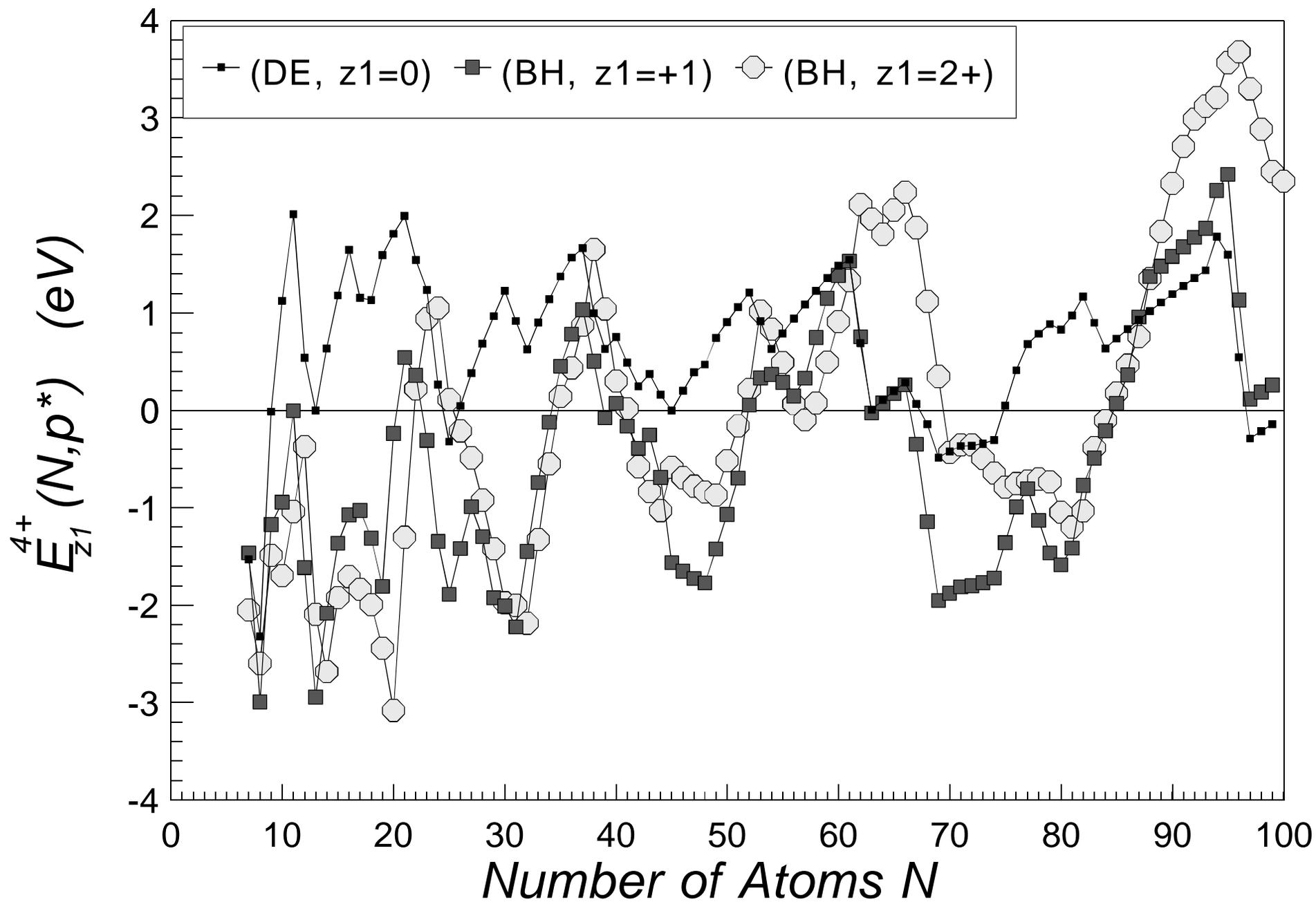
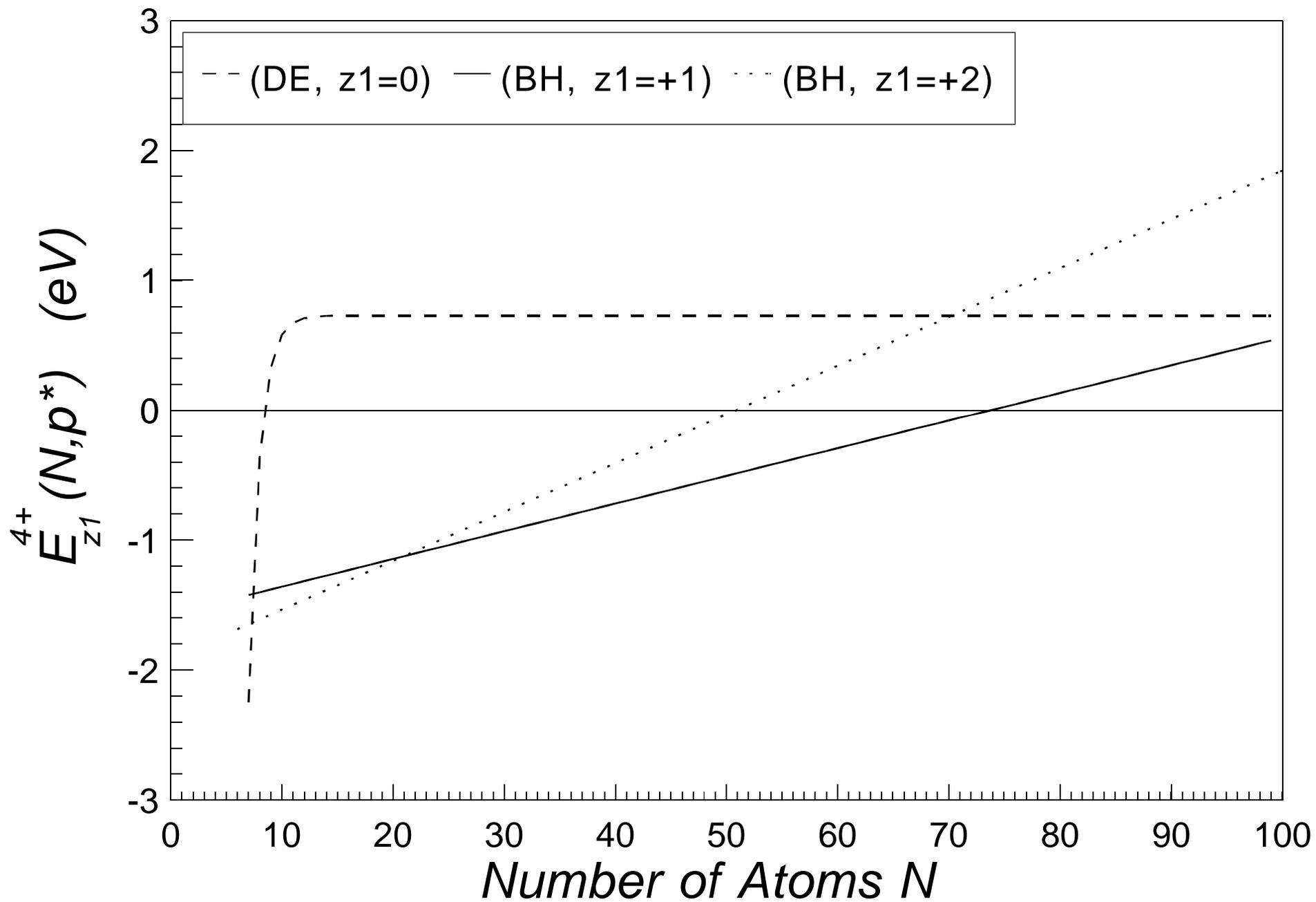
\begin{figure}
\caption{(a)- The same as in Fig. \protect\ref{fig4}(a) for 4-ply ionized
clusters, with the asymptotic value of the dashed fitted line as $0.75 eV$.
(b)- The same as in Fig. \protect\ref{fig5}(b) for charge-asymmetric
fission with the intersection point at $N_1=74$.
(c)- The same as in Fig. \protect\ref{fig5}(b) for charge-symmetric
fission with the intersection point at $N_2=51$.
(d)- Comparison of the evaporation DE's with the charge-asymmetric ($z_1=1$)
and the charge-symmetric ($z_1=2$) fission BH's.
(e)- Comparison of the fitted lines for the processes in (d). At $N=20$, the
charge-symmetric and the charge-asymmetric fissions compete. For $N>51$,
charge-symmetric and for $N>73$, the charge-asymmetric fissions stop,
respectively. For $N_c^{\rm mean}>107$ evaporation overcomes the fission
processes.}
\label{fig7}
\end{figure}

Fig. \ref{fig7}(a) plots the $p^*$ and the $D^{4+}_0(N,p^*)$ for the
evaporation processes of Ag$_N^{4+}$.
In this figure, one notes the unstability of two smallest sized
Ag$_7^{4+}$, Ag$_8^{4+}$ and the stability of almost all others against the
evaporation. The fitted dashed line predicts no evaporation
and has
the  asymptotic value of $0.75 eV$. The evaporation
products  are seen to be mostly neutral dimers and a few monomers and octamers.

 In Fig. \ref{fig7}(b) we have plotted the $p^*$ and the
 $B^{4+}_{1+}(N,p^*)$ as functions of $N$. We see that at $N=21$ the BH becomes
positive for the first time.  The dashed line shows the mean behavior of the
fission barries.
This fitted line has crossed the zero axis at $N_1^{\rm mean}\approx 74$. That
is, on the average, the Ag$_N^{4+}$ clusters are stable against the
charge-asymmetric fission channel for $N>74$. The most favored charged products
are mostly magic clusters, Ag$_N^{1+}$ with $N=3,9,21$.

The results of charge-symmetric binary fission of Ag$_N^{4+}$ for most favored
decays are shown
in Fig. \ref{fig7}(c). The smallest positive BH occurs at $N=22$. The mean
behavior
(the dashed line) intersects  the zero line at $N_2^{\rm mean}\approx 51$.

In Fig. \ref{fig7}(d) we have compared the results for the three different
decay processes of Ag$_N^{4+}$. It is seen that these clusters smaller than
$N\approx 50$ are highly unstable.
As is seen from the figure, it is difficult to specify the
competitions for the quantal values. However, to give quantitative values, we
have compared the fitted lines in
Fig. \ref{fig7}(e). It is seen that for $N<20$ the charge-symmetric fission is
the dominant spontaneous decay process but, for $20<N<51$ the dominant
spontaneous fission process changes to the charge-asymmetric one. For $51<N<73$
the only spontaneous fission decay process is the charge-asymmetric one.
Clusters larger than $N_a^{\rm mean}\approx 73$ are stable against any
spontaneous decay process.
In an induced fragmentation experiment of Ag$_N^{4+}$ clusters, the dominant
process for $73<N<107$ is charge-asymmetric fission, and for $N$ larger than
$N_c^{\rm mean}=107$ the evaporation process dominates. To summarize, for
smaller
clusters the charge-symmetric fission is dominant, and larger clusters prefer
to decay via a charge-asymmetric fission process.

Besides the most favored quantities which are strongly related to the
stability of the charged cluster and were explained in the above lines, it
is also interesting to calculate the DE's and BH's for a process in
which the fragment products are specified.
Consider the process
${\rm Ag}_N^{Z}\to {\rm Ag}_{N-1}^{Z} + {\rm Ag}^{0}$ in which one of the
products is a neutral monomer. We have calculated the dissociation energies
$D^Z_0(N,1)$ for all values of $Z$=1, 2, 3, 4 and $N\le 100$. The calculated
values show pronounced shell effects as in previous figures of the most
favored channels.
However, the mean behaviors have asymptotic constant values. For $Z=1,2,3,4$,
these asymptotic values in electron-volts are 0.95, 1.00, 1.08, 1.15,
respectively. This means that the monomer evaporation from a
singly charged cluster needs a smaller energy than from a doubly charged and so
on. The same analysis for the dimer evaporation in the process
${\rm Ag}_N^{Z}\to {\rm Ag}_{N-2}^{Z} + {\rm Ag}_2^{0}$ shows also constant
asymptotic mean behaviors for the $D^Z_0(N,2)$. The obtained values in
electron-volts are 0.49, 0.59, 0.72, 0.87 for $Z$=1, 2, 3, 4, respectively.
This means that, as in the monomer evaporation, the detachment of a dimer from
singly ionized cluster is easier than from a doubly ionized cluster and so on.
However, comparing the dissociation energies for monomer and dimer evaporation
(keeping the charge constant) shows that atomic evaporation needs more energy
than dimer evaporation.

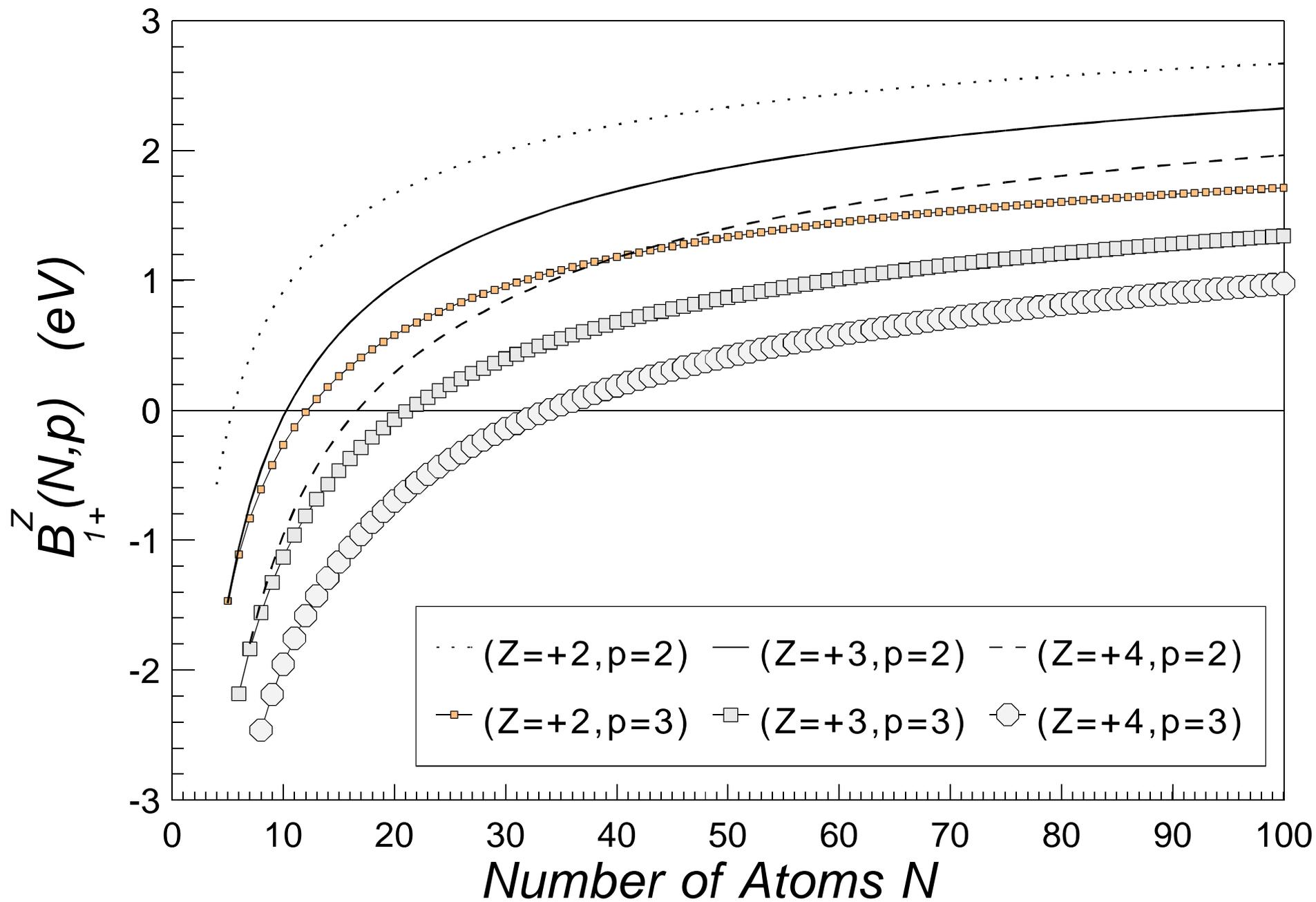
\begin{figure}
\caption{Comparison of the fitted values of the BH's for dissociating a singly
ionized dimer or trimer from a $Z$-ply charged cluster.}
\label{fig8}
\end{figure}

Now, we consider the fission processes
${\rm Ag}_N^{Z}\to {\rm Ag}_{N-2}^{Z-1} + {\rm Ag}_2^{1+}$ and
${\rm Ag}_N^{Z}\to {\rm Ag}_{N-3}^{Z-1} + {\rm Ag}_3^{1+}$
in which one of the fission products is a singly ionized dimer or a
singly ionized trimer. The mean behaviors of the BH's for these
processes are plotted in Fig. \ref{fig8}. As is seen, the energy needed to
detach a singly ionized dimer decreases by increasing the charge of the
parent cluster. This behavior should be contrasted to the behavior in the
monomer or dimer evaporations. We recall that in the monomer or dimer
evaporation, the dissociation energy increases by increasing the charge of the
parent cluster.

It is now easy to find the mean sizes at which atomic evaporation process
dominates
the fission into singly ionized dimer or trimer for each charging value of the
parent cluster. In doubly charged silver clusters, the monomer evaporation
dominates the singly charged dimer and trimer detachments at $N=11$ and $N=31$,
respectively. The corresponding numbers for triply charged clusters are 21 and
66. For parent clusters Ag$_N^{4+}$, the numbers $N=38$ and $N=120$ have been
obtained. To summarize, by increasing the charge of the parent cluster the
competition occurs at larger values of $N$.

\section{Conclusion}
\label{sec4}

In this work, we have studied the fragmentations of multiply charged silver
clusters taking into account the structural relaxations of the neutral and
charged parent as well as daughter clusters. To calculate the relaxed-state
sizes and energies of the clusters we have employed the stabilized jellium
model with self-compression using a spherical geometry for the jellium
background. Using these relaxed-state radius and energy for the clusters, we
have calculated the dissociation energies and barrier heights for evaporation
and fission processes in all possible channels. For the barrier heights, we
have used the two-touching-spheres model with taking into account the
polarizabilities of the two charged products. Comparison of our most
favored results with
the experimental data shows that our results lie under the experimetal results
but, the critical size for the competition of the evaporation and the fission
of doubly
charged silver clusters is predicted in good agreement with the experiment.
This comparison also reveals that the fragmentation processes mostly occur
before the complete relaxation of the charged parent clusters. That is, in the
above-mentioned experiments the structural relaxation time is larger than the
average time elapsed for the fragmentation of the ionized parent cluster.
Having
the initial (just after ionization) and the relaxed sizes $r_{s,\rm ini}^Z$,
$r_{s,\rm rela}^Z$ of a $Z$-ply ionized cluster,
one may choose an $r_{s,\rm frag}^Z$ value
($r_{s,\rm ini}^Z\le r_{s,\rm frag}^Z\le r_{s,\rm rela}^Z$)
for the ionized cluster (just before the
fragmentation) such that the calculated values coincides the experimental ones.
Then using a linear interpolation it is possible to calculate the relative
fragmentation time for an ionized cluster. In ordinary jellium model
calculations, the assumption

\begin{equation}
r_s^0(N)=r_s^Z(N)=
r_{s,\rm ini}^Z(N)=r_{s,\rm frag}^Z(N)=r_{s,\rm rela}^Z(N)=r_{s,\rm bulk}
\label{eq29}
\end{equation}
is used. It should be mentioned that for exact matching of the calculated and
experimental values one should use non-spherical shapes.

We have obtained the asymptotic DE's for the most favored channels in
evaporation processes by fitting
a simple curve on the quantal results. The result shows that the asymptotic
values increase by increasing the charge of the parent cluster. In the case of
Ag$_N^{4+}$, we have shown that for relatively small clusters the
charge-symmetric fission process is dominant and then, before dominating the
evaporation process the charge-asymmetric fission process overcomes.
In general, the critical size (at which the evaporation dominates the
fission) increases by increasing the charge of the parent cluster. The results
show that the neutral $p$-mer dissociation energy increases by increasing
the charge of the parent cluster; and for a given charged parent, the atomic
evaporation needs more energy than a dimer evaporation. Finally, it has been
shown that the energy needed for the detachment of a singly charged dimer or
singly charged trimer decreases by increasing the charge of the parent cluster.
However, for a given parent cluster, the detachment of a singly charged trimer
is easier than that of a singly charged dimer.

{\large\bf Acknowledgement}

{The author would like to thank John P. Perdew for the useful discussions
and comments during this work. He also thanks Adam Kiejna for providing me with
his recent review article on the stabilized jellium model.}

\newpage

\end{document}